\documentclass[usenatbib, usegraphicx]{mn2e}
\usepackage{amssymb}
\usepackage{amsmath}
\def\aj{AJ}
\def\apjs{ApJS}
\def\mnras{MNRAS}
\def\araa{ARA\&A }
\def\nat{Nature}

\def\apj{ApJ}
\def\aap{A\&A} 
\def\apjl{ApJL}
\def\aaps{A\&AS}
\title[Quasar selection in VIKING]{Selection constraints on high redshift quasar searches in the VISTA kilo-degree infrared galaxy survey}

\author[J.~R.~Findlay et al.]
{\parbox{\textwidth}{J.~R.~Findlay,$^{1}$\thanks{E-mail: \texttt{j.findlay@qmul.ac.uk}}
W.~J.~Sutherland,$^{1}$
B.~P.~Venemans,$^{2}$
C.~Reyl\'e,$^{3}$
A.~C.~Robin,$^{4}$
D.~G.~Bonfield,$^{4}$
V.~A.~Bruce$^{5}$ and
M.~J.~Jarvis.$^{4,6}$}\vspace{0.4cm}\\
\parbox{\textwidth}{$^{1}$Astronomy Unit, Queen Mary, University of London, London E1 4NS.\\
$^{2}$European Southern Observatory, Karl-Schwarzschild-Strasse 2, 85748 Garching bei Munchen,
  Germany.\\
$^{3}$Universit\'e de Franche-Comt\'e, Institut UTINAM, UMR CNRS 6213, Observatoire des Sciences de l'Univers THETA de Franche-Comt\'e, Observatoire de Besan\c{c}on, BP 1615, 25010 Besan\c{c}on Cedex, France.\\
$^{4}$Centre for Astrophysics, Science \& Technology Research Institute, University of Hertfordshire, Hatfield, Herts, AL10 9AB, UK.\\
$^{5}$SUPA Institute for Astronomy, University of Edinburgh, Royal Observatory, Edinburgh EH9 3HJ\\
$^{6}$Physics Department, University of the Western Cape, Cape Town, 7535, South Africa.\\
}}

\begin{document}

\date{Accepted XXX. Received XXX; in original form XXX}

\pagerange{\pageref{firstpage}--\pageref{lastpage}} \pubyear{XXX}

\maketitle

\label{firstpage}

\begin{abstract}

  The European Southern Observatory's (ESO) Visible and Infrared Survey Telescope for Astronomy (VISTA) is a 
  4-m class survey telescope for wide-field near-infrared imaging. VISTA is currently running a suite 
  of six public surveys, which will shortly deliver their first Europe wide public data releases to ESO. 
  The VISTA Kilo-degree Infrared Galaxy Survey (VIKING) forms a natural intermediate 
  between current wide shallow, and deeper more concentrated surveys, by targeting two patches totalling 
  $1500\,\mathrm{deg}^2$ in the northern and southern hemispheres with measured 5$\sigma$ limiting depths of 
  ${\rm Z} \simeq 22.4, {\rm Y} \simeq 21.4, {\rm J} \simeq 20.9, {\rm H} \simeq 19.9~\mathrm{and}~
  {\rm Ks} \simeq 19.3$ (Vega). This architecture forms an ideal working parameter space for the discovery of
  a significant sample of $6.5 \leq \mathrm{z} \leq 7.5$ quasars. In the first data release priority has been placed on 
  small areas encompassing a number of fields well sampled at many wavelengths, thereby optimising science gains 
  and synergy whilst ensuring a timely release of the first products. For rare object searches e.g. high-z 
  quasars, this policy is not ideal since photometric selection strategies generally evolve considerably
  with the acquisition of data. 
  Without a reasonably representative data set sampling many directions on the sky it is not clear
  how a rare object search can be conducted in a highly complete and efficient manner.

  In this paper, we alleviate this problem by supplementing initial data with a realistic model 
  of the spatial, luminosity and colour distributions of sources known to heavily contaminate photometric
  quasar selection spaces, namely dwarf stars of spectral type M, L and T. We use this model along with 
  a subset of available data to investigate contamination of quasar selection space by cool stars and galaxies
  and lay down a set of benchmark selection constraints that limit contamination to reasonable levels whilst 
  maintaining high completeness as a function of both magnitude and redshift. We review recent
  follow-up imaging of the first VIKING high-z quasar candidates and find that the results lend 
  considerable support for the choice of selection constraints. The methods outlined here are 
  also applicable to rare object searches in a number of other ongoing and forthcoming projects.

\end{abstract}


\section{Introduction}\label{sec:intro}
The existence of luminous quasars at ${\rm z} \sim 6$\footnote{There are a number
of variables which are conventionally denoted z or Z. To avoid confusion they are 
defined here as z; redshift, $z$; the SDSS passband, 
Z; the VISTA passband.} 
suggests that the formation of the first supermassive black holes began within the first 
few hundred million years of the big bang. As such high-z quasars offer 
powerful probes of the early Universe, setting constraints on early 
structure formation \citep[][etc.]{2005ApJ...622L...1S, 2005ApJ...626..657W,2007AJ....133.2222S}, 
chemical enrichment \citep[][and refs.~therein]{2009AIPC.1111..160M} and 
the state of the inter-galactic medium at the end of cosmic re-ionisation 
\citep{2006AJ....132..117F}.

In the last decade, optical surveys such as the Sloan Digital Sky 
Survey \citep[SDSS;][]{2000AJ....120.1579Y} and the Canada France High-z 
Quasar Survey \cite[CFHQS;][]{2005ApJ...633..630W} have 
increased the number of known quasars at $\mathrm{z} \sim 6$ to over 
50 \citep[e.g.][]{2008AJ....135.1057J,2010AJ....139..906W}. At the single scan sensitivity limit of the Sloan camera 
($z \simeq 21.0$) the surface density of $\mathrm{z} \geq 6$ quasars brighter
than $z \leq 21$ is of order $\rho_\mathrm{Q} \simeq 0.01\,\mathrm{deg}^{-2}$,
so this represents a considerable effort.

High-z quasars are generally initially selected for follow-up 
observations by virtue of their photometric colours.
The key to the photometric detection of a 
strong quasar candidate is to isolate the characteristic flux break 
blueward of the Lyman-alpha (Ly-$\alpha$) transition brought about by the thickening 
of the Ly-$\alpha$ forest towards higher redshifts. Colour selection 
techniques exploit this break by placing a blue observing band across the 
Ly-$\alpha$ transition and a further band just redward. Objects suitable for 
follow-up spectroscopy 
should then present themselves as extremely red faint point sources. A 
carefully placed colour cut is usually sufficient to confine the most 
interesting sources to a region of colour space distinct from most main 
sequence stars and galaxies. In principle, this facilitates candidate 
selection in just two passbands, but in practice a third passband is 
necessary to permit a measurement of the continuum level redward of 
Ly-$\alpha$ as a means of reducing contamination from cool degenerate 
stars. This is the technique pioneered by the SDSS and adopted in the 
CFHQS; both make initial selections in $i-z$ and take follow-up 
${\rm J}$-band imaging of promising candidates prior to spectroscopic confirmation
\citep{2001AJ....122.2833F,2005ApJ...633..630W}.

At $\mathrm{z} > 6$ Ly-$\alpha$ begins to shift out of the Sloan 
$z$ band and while projects like the Panoramic Survey Telescope and 
Rapid Response System \cite[Pan-STARRS;][]{2002SPIE.4836..154K} aim to progress towards higher redshifts
in a predominantly optical parameter space (by selecting $i$ dropouts), faint optical
detections make it difficult 
to reject the numerous Galactic stars with scattered quasar-like 
colours that can out number high-z quasars by a factor $10^4$.

To this end a number of other surveys have taken a different approach, 
employing the near-IR as their selection space, a tactic which has had 
recent encouraging success in the UKIRT Infrared Deep Sky Survey \cite[UKIDSS;][]{2007MNRAS.379.1599L}. 
One of the innovations behind the UKIDSS photometric system, was to 
recognise that near-IR filter combinations incorporating $z$, ${\rm J}$, ${\rm H}$ and ${\rm K}$ alone 
are inadequate for $\mathrm{z} > 6$ quasar selection, since cool 
star spectra tend to peak in the near-IR and crowd the quasar colour-colour 
locus \citep{2002ASPC..283..369W}. The solution was to introduce an
observing band intermediate between $z$ and ${\rm J}$ and was put forward by 
\citet{2006MNRAS.367..454H}. The ${\rm Y}$-band filter was optimised for this purpose 
and has been used to great effect in the UKIDSS-LAS 
\citep[UKIDSS Large Area Survey;][]{2007MNRAS.379.1599L} where seven ${\rm z} \gtrsim 6$ quasars have been identified  
up to the UKIDSS eighth data release 
\citep[][Venemans~et.~al.\ in preparation, Patel~et~al.\ in preparation]{2007MNRAS.376L..76V,2009A&A...505...97M,2011Natur.474..616M}.       

Working in the same parameter space as UKIDSS is the European Southern Observatory's (ESO)
Visible and Infrared Survey Telescope for Astronomy (VISTA), a 4-m class wide field 
survey telescope located at ESO's Paranal observatory in Chile \citep{2010Msngr.139....2E}. 
More than 80 per cent of VISTA time is devoted to 
running a suite of six public surveys, which will deliver European public data releases
in the near future. Of these projects, the VISTA Kilo-degree Infrared 
Galaxy survey (VIKING) has a combination of depth and area that is ideally 
suited for quasar searches at the highest redshifts. Over the nominal five 
year survey life span, VIKING will image a total of $1500\,\mathrm{deg}^2$ 
in the ${\rm Z}$, ${\rm Y}$, ${\rm J}$, ${\rm H}$, ${\rm Ks}$ near-IR bandpasses 
centred on the northern Galactic cap and the 
southern Galactic pole. VIKING will detect in excess of $6 \times 10^6$ stars and 
$20 \times 10^6$ galaxies down to a limiting $5\sigma$ detection limit 
of ${\rm J} \simeq 20.9$ (Vega system).

\begin{figure*}
  \advance\leftskip-0.8cm
\includegraphics[scale=0.49]{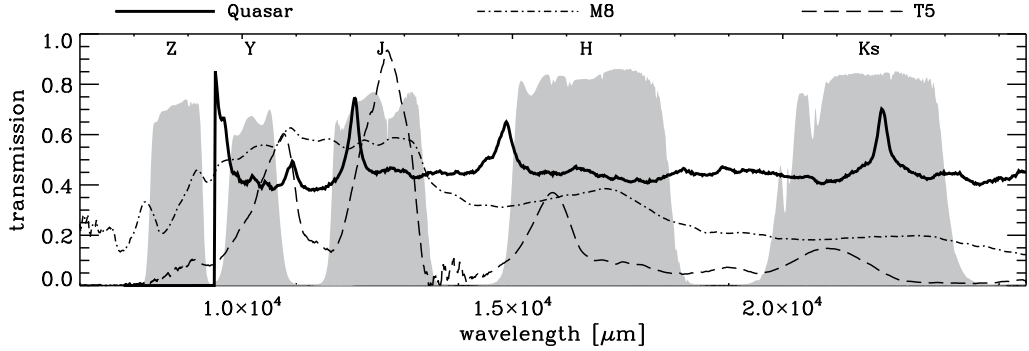}
\caption{Transmission curves for the VISTA ${\rm Z}$, ${\rm Y}$, ${\rm J}$, ${\rm H}$, ${\rm Ks}$ 
  photometric system, including
  wavelength dependent losses are labelled and shaded grey. Also shown is the LBQS
  template quasar spectrum artificially redshifted to $\mathrm{z} = 6.8$ and with
  the characteristic absorption trough imposed blueward of redshifted Ly-$\alpha$
  (solid black curve). For comparison the spectra of cool M8- and T5-dwarfs are also plotted 
  (dash-dot and dashed curves respectively).}
\label{fig:filters}
\end{figure*}

VISTA's transmission curves are shown in Figure~\ref{fig:filters}.
These include the relevant contributions from the instrument and detector, 
mirror reflectivity and transmissions through the atmosphere and 
filters under typical photometric conditions. Also plotted is the Large 
Bright Quasar Survey's \citep[LBQS;][]{1995AJ....109.1498H} template quasar spectrum 
\citep{1991ApJ...373..465F}, redshifted to $\mathrm{z} = 6.8$ and with the characteristic 
HI absorption trough artificially imposed blueward of redshifted Ly-$\alpha$ at 
\mbox{$\lambda\,($Ly-$\alpha) = 1216\, \mathrm{\AA} \times 7.8$}. The spectra of cool 
M8- and T5-dwarf stars are plotted alongside the quasar spectrum to highlight,
the recognised difficulty of differentiating between high-z 
quasars and cool stars at the foot of the main sequence and below 
the hydrogen burning cut-off.

M-dwarfs out-number high-z quasars by at least a factor $\sim 10^5$ 
(e.g.\ \citet[][]{2010AJ....139.2679B}, obtained a volume limited sample of 
$\sim 15 \times 10^6$ SDSS objects with M-dwarf like colours at $z < 21.2$.
Over a similar depth and area there are currently 19 SDSS quasars with ${\rm z} \sim 6$
reported by Fan et.\ al.\ 2006.), while the cooler L and T spectral types are by all
accounts, a lot more rare \citep{2009MNRAS.397..258L,2010MNRAS.406.1885B}. 
The broad similarities between high-z quasar and M-dwarf spectra,
coupled with their relative surface densities means that most colour selected high-z quasar
candidates are in fact M-dwarfs with scattered quasar like colours resulting from large photometric 
errors. The function of a good colour selection criterion then, is to maintain a large selection space
while simultaneously minimising contamination from false positives and optimising the 
completeness of the search. 

In the vein of other wide field imaging surveys before it, the initial VIKING 
data release consists of a few high priority fields to provide a testbed
for future larger releases. In the first public VIKING data release  
$\sim75\,\mathrm{deg}^2$ will be made available.
Precision selection strategies can be developed with a 
detailed understanding of the error perturbed magnitude and colour distributions of the 
target and contaminant populations, both as a function of redshift and direction on the sky. 
These strategies will no doubt develop with the accumulation of large amounts of data. However, 
the first photometric follow-up of VIKING candidates was undertaken in summer 2011. 
Although by this time there was $\sim 350\,\mathrm{deg}^2$ of data available, there was 
insufficient time to analyse it in detail other than to select candidates based on selection 
constraints derived from smaller initial data sets. 
Thus, to optimise selection constraints and provide a complete, efficient and
non-biased search it was necessary to use a combination of initial data and 
detailed modelling \citep[e.g.][]{1999AJ....117.2528F}.  

The following article describes a model of the magnitude, colour and spatial
distributions of the cool star and quasar populations, which was developed
and assessed along with  $\sim \,200\, \mathrm{deg}^2$ of initial imaging
data to lay down some baseline magnitude and colour selection criteria to 
apply to the $\sim \, \mathrm{350\, deg}^2$ available for the first follow-up
observations.

The paper is organised as follows; In Section~\ref{sec:synphot} VIKING Vega system
synthetic photometry is computed for typical examples of cool-stars and quasars. In Section~\ref{sec:offset} 
offsets from the Vega system are measured in the VIKING data.
In Section~\ref{sec:model} the synthetic photometry is employed along with 
number count models to populate a synthetic realisation of the cool-star and high-z
quasar component of the VIKING catalogue. In Section~\ref{sec:selection} the
catalogue and available data are used to place constraints on quasar selection space and assess 
completeness. In Section~\ref{sec:ntt} we assess the selection criterion in light of recent
followup observations. We summarise our findings and outline future work in 
Section~\ref{sec:sum}.

Henceforth, all quoted photometry is defined on the Vega system,
such that Vega is a zero magnitude star in all \mbox{VISTA} passbands. Magnitudes
are quoted in the traditional logarithmic form except for those in the
${\rm Z}$ band, which are quoted in the $\mathrm{asinh}$ form 
\citep{1999AJ....118.1406L}. $\mathrm{asinh}$ magnitudes take the same values as the 
equivalent logarithmic form for high signal-to-noise (S/N) measurements, but have the
advantage of linear dependence at low S/N or even negative flux,
which will be the case for most $\mathrm{z} \gtrsim 6.5$ quasars. A softening
parameter denoted $b$, is required to define the scale at which linear dependence
dominates. For the ${\rm Z}$ band we choose $b$ 
equivalent to ${\rm Z} = 24.1$, which
gives the $\mathrm{asinh}$ magnitude of an object with zero flux and is set
approximately equal to the sky noise seen in the initial data frames.
 
Throughout this work we adopt the following values for the
Hubble constant and the matter and cosmological constant energy
density parameters; 
$H_0 = 70\,{\rm km}\,{\rm s}^{-1}\,{\rm Mpc}^{-1}$, 
$\Omega_{M} = 0.28$ and 
$\Omega_{\Lambda} = 0.72$ respectively \citep{2009ApJS..180..330K}.

\section{Z, Y, J Synthetic Photometry and Characteristic Colours}\label{sec:synphot}

Characteristic colours were calculated for a set of cool stars and quasars by integrating the product
of their spectra and the VISTA response
functions\footnote{http://www.eso.org/sci/facilities/paranal/instruments/vista/inst/}, including 
measured values for the wavelength--dependent losses  resulting from atmospheric extinction, 
mirror reflectivity and instrument and detector response.

\subsection{High-redshift quasars}

In order to obtain a representative spread of quasars in
 colour--colour space, we use a sample of 386 quasars from
 the \mbox{SDSS} Quasar Catalogue~IV \citep{2007AJ....134..102S}. The
 sample spans a redshift range of \mbox{$3.1 \lesssim \mathrm{z} \lesssim 3.2$}, over
 which there is low stellar contamination and
 high completeness \citep{2002AJ....123.2945R}. 

 Recent comparisons between high and low-z quasar spectra have shown
 no obvious evidence for metalicity evolution in the rest-frame UV spectral
 energy distributions \citep[][although see Jiang et al.\ 2010 for possible 
evolution of dust properties]{2007AJ....134.1150J, 2009A&A...494L..25J,2011Natur.474..616M}. 
Thus, it is a reasonable approximation to simulate intrinsic
 quasar spectral energy distributions at \mbox{$\mathrm{z} \ge 6.5$} by artificially redshifting the low-z SDSS sample. 

Significant differences between  \mbox{$\mathrm{z} \gtrsim 6.5$} quasars and those 
at lower redshift are apparent blueward of \mbox{Ly-$\alpha$} due to
 increasingly strong absorption in the intergalactic medium with redshift. 
At \mbox{$\mathrm{z} = 6.5$} the optical depth of the intergalactic medium to 
 \mbox{Ly-$\alpha$} photons is $\gg$~1 \citep[e.g.][]{2006AJ....132..117F}. 
 For the same neutral density the optical depths to \mbox{Ly-$\beta$} and 
 \mbox{Ly-$\gamma$} are factors of 6.2 and 17.9 less than for that
  of \mbox{Ly-$\alpha$}, but their line strengths are comparatively small and they
  can be reasonably neglected \citep[e.g.~][]{1966atp..book.....W}. 
  Transmission blueward of \mbox{$\lambda = 1216 \, \mathrm{\AA} \,(1+\mathrm{z})$}  
 is thus effectively nil, and absorption by intergalactic hydrogen can
  be adequately modelled in \mbox{$\mathrm{z} \geq 6.5$} quasar spectra by
  setting transmission shortward of redshifted \mbox{Ly-$\alpha$} to zero.

\subsection{Stars}

A library of cool star spectra has been collected from the literature 
 covering \mbox{VISTA's} \mbox{Z, Y, J, H and ${\rm K_s}$} passbands, 
 these include all 
 \mbox{21 M-}, \mbox{30 L-} and \mbox{22 T-dwarf} spectra on which \citet{2006MNRAS.367..454H} 
conducted synthetic photometry
  in the \mbox{WFCAM} passbands. References for these objects
  can be found in the above paper. 
 The library is further supplemented by spectra obtained from the 
 DwarfArchives\footnote{http://spider.ipac.caltech.edu/staff/davy/ARCHIVE/links.shtml},
 the IRTF (NASA Infrared Telescope Facility) spectral library\footnote{http://irtfweb.ifa.hawaii.edu/$\sim$spex/IRTF\_Spectral\_Library/}, 
 the SpeX Prism spectral library\footnote{http://pono.ucsd.edu/$\sim$adam/browndwarfs/spexprism/},
 the Keck LRIS (Low Resolution Imaging Spectrometer) spectral library\footnote{http://www.stsci.edu/$\sim$inr/ultracool.html} and 
 Sandy Leggett's L- and T-dwarf archives\footnote{http://staff.gemini.edu/$\sim$sleggett/LTdata.html}.

\subsection{Selection in Z, Y, J}\label{ssec:zyj}

The resulting synthetic colours have been checked for consistency
by convolving stellar sources in the \mbox{WFCAM} passbands and comparing
results with those presented for the same sources by 
\citet{2006MNRAS.367..454H} finding good agreement between the two. 
There is also broad agreement between these results and 
observational work 
\citep[e.g.][]{2009A&A...497..619Z, 2010ApJ...710.1627L}. 
Simulated \mbox{$i - z$} colours of the quasar sample over
the \mbox{SDSS} passbands are also consistent with the corresponding 
\mbox{SDSS} photometry.  

\begin{figure}
\advance\leftskip-0.1cm
\includegraphics[scale=0.5]{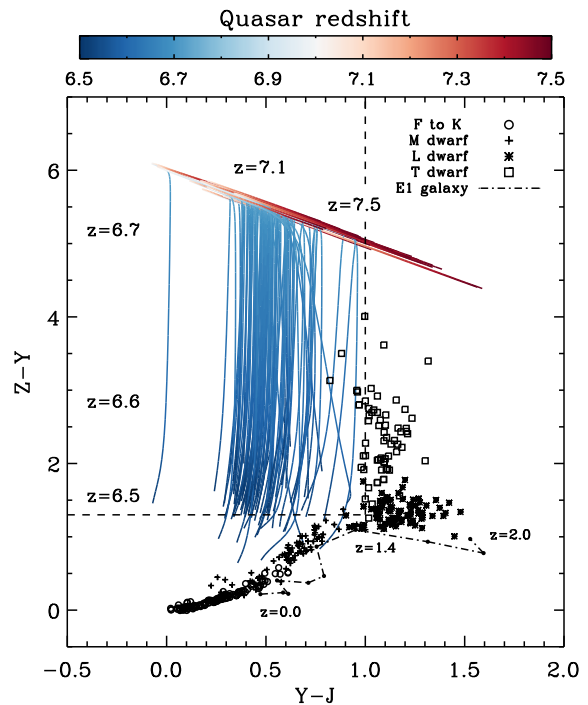}
\caption{The ${\rm Z}$, ${\rm Y}$, ${\rm J}$ colour plane; high-z quasar redshift evolution tracks are colour coded
  as indicated by the colour bar. Star and galaxy colours are indicated by the legend. The
  dashed line describes the general principles of quasar colour selection in the ${\rm Z}$, ${\rm
  Y}$, ${\rm J}$ passbands,
  enclosing the vast majority of high-z quasars while rejecting the vast majority of galaxies and
  foreground stars.}
\label{fig:zyj}
\end{figure}

At this stage it is useful to review the general approach to near-IR high-z quasar
selection in visual form. In Figure~\ref{fig:zyj} a ${\rm Z}$, ${\rm Y}$, ${\rm J}$ colour-colour diagram is presented
with the derived synthetic photometry. The quasar tracks are colour coded 
according to their redshift, in steps of \mbox{$\Delta z = 0.01$} and the average
track is labelled at various redshifts for clarity. The box 
bounded by the dashed line illustrates the general principle;
the box includes almost all \mbox{$6.5 \leq \mathrm{z} \leq 7.5$} quasars, 
rejecting galaxies as shown by the E1 locus \citep{2008MNRAS.386..697R} and almost 
all foreground stars.

 The position of each quasar in \mbox{${\rm Z-Y}$} is a function of redshift, reflecting the movement of
 the Ly-$\alpha$ break through the Z-band. The \mbox{$\rm{Y-J}$} colours have only minor
 dependence on redshift until \mbox{$\mathrm{z} = 6.9$} where the locus begins to
 turn over as \mbox{Ly-$\alpha$} enters the observed ${\rm Y}$ band. Conversely, the distribution 
 of stellar colours is driven largely by
 temperature, which acts to redden spectral types approaching early-L. 
 In later types, behaviour is more 
 complicated as clouds act to mask spectral features and redden
 near-IR colours \citep[see][]{2005ARA&A..43..195K}. 
 With the appearance of pressure broadened NaI and KI doublets in the reddest optical bands 
 at mid to late L, the sequence is forced redward in \mbox{${\rm Z-Y}$} into the T sequence.
 
 The simple cuts outlined above will need to be modified
 for a real survey, since random photometric errors will
 broaden the measured locus, and thus can scatter
 stars and galaxies into the quasar selection region. Since all stars and galaxies
 outnumber genuine high-z quasars by a factor \mbox{$> 10^6$} $({\rm J} \lesssim 21)$, a realistic 
 survey will need to ensure that the cuts are sufficiently far from the
 stellar locus to reject almost all stars and galaxies. In the reminder of this
 article we examine how best to place these cuts such that they isolate the
 vast majority of high-z quasars without selecting unacceptable numbers of 
 contaminants.
 
\section{Z, Y Offsets from the Vega System}\label{sec:offset}

VISTA photometry is calibrated with reference to a set of 2MASS 
\cite[the Two Micron All-Sky Survey;][]{2006AJ....131.1163S} standards,
which have been measured with reference to the spectrophotometric
standard star Vega ($\alpha\mathrm{-Lyrae}$). This implies that
Vega is a zeroth magnitude star in all VISTA bandpasses. The calibration
involves measuring the offsets between the 2MASS standards and
VISTA observed stars via a set of colour equations. Since 2MASS photometry was
measured in the ${\rm J}$, ${\rm H}$, ${\rm Ks}$ bands alone the basic assumption is that the
2MASS colour equations can be linearly extrapolated to cover the VISTA 
${\rm Z}$ and ${\rm Y}$ bandpasses. Any divergence from linearity will result in an offset
from the Vega system in the extrapolated wavelength region.

A similar occurrence was noted by \citet{2009MNRAS.394..675H} in UKIDSS data release 2, where a 
rigorous analysis of the ${\rm Z}$, ${\rm Y}$, ${\rm J}$, ${\rm H}$ and ${\rm K}$ photometry 
against the overlapping
SDSS footprint found a significant offset in the ${\rm Y}$-band. A similar
analysis for VISTA will no doubt be attempted when the catalogues are 
diverse enough to make precise measurements. However, since our
quasar selection criterion is to be partly based on Vega zeroed 
synthetic photometry, it is important to at least tentatively determine any
offsets present before we begin.

VIKING data is processed by the VISTA Data Flow System \citep{2004SPIE.5493..401E} in 
its pipeline \citep{2004SPIE.5493..411I} and retrieved from the VISTA Science Archive\footnote{
The VSA holds the image and catalogue data products generated 
by the VISTA Infrared Camera (VIRCAM). 
The primary contents of the archive originates from the VISTA Public Surveys.
The archive can be queried via the Structured Query Language (SQL) and several
interactive web forms at http://horus.roe.ac.uk/vsa/index.html} \citep[VSA;][]{2004SPIE.5493..423H}.
At the time of this analysis the VSA contains the latest v1.0 VIKING release which comprises of some 
$\mathrm{200\,deg}^2$ of imaging in ${\rm Z}$, ${\rm Y}$, ${\rm J}$ and $\mathrm{120\,deg}^2$
of imaging in all five bands; $90\,\mathrm{deg}^2$ of this has 
complementary SDSS overlap. Offsets in this data set are measured following the
approach of \citet{2009MNRAS.394..675H}, the main points of which
are summarised here. The reader should refer to the original paper
for a full description including the maximum likelihood fitting procedures
referred to below.

\begin{figure}
\begin{center}
\advance\leftskip-0.8cm
\includegraphics[scale=0.4]{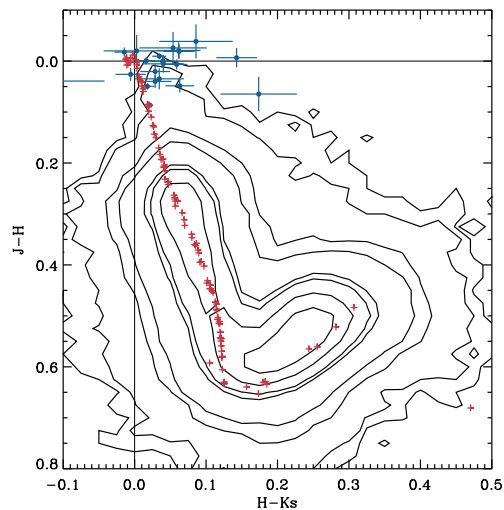}
\caption{The VIKING ${\rm J}$, ${\rm H}$, ${\rm Ks}$ colour plane for high confidence stellar sources matched in VIKING
  and SDSS. Blue points are stars with Vega like colours in all SDSS passband combinations. Red
  crosses are synthetic photometry from BPGS spectroscopic atlas \citep{1983ApJS...52..121G}.}
\label{fig:jhks}
\end{center}
\end{figure}

A sub-sample of high signal-to-noise (S/N $\geq 10.0$) point-like sources detected in both 
VIKING and the SDSS was defined and those sources with Vega like SDSS colours i.e. 
A0 stars with $u-g$, $g-r$, $r-i$ and $i-z$ in the range -0.1 to 0.1, were flagged.
Figure~\ref{fig:jhks} plots the ${\rm J}$, ${\rm H}$ and ${\rm Ks}$ colour-colour diagram for this set of sources. Blue points
show the photometrically selected sample of A0 stars. The small number of these objects
highlights the difficulty of investigating the VIKING colours at this early stage.
The median colours of this sample are $\mathrm{J}-\mathrm{H}=0.005\,\pm \,0.038$ and 
$\mathrm{H}-\mathrm{Ks}= 0.033\,\pm \,0.023$ (where the median absolute deviation has been used to
estimate the standard deviation), which perhaps hints at
small offsets in one or both of ${\rm H}$/${\rm Ks}$. Given that each of these colours
has been directly calibrated from 2MASS standards, we continue our analysis
assuming that any offset in these passbands is small. As we will show shortly,
there is evidence that this assumption is a good one. Also plotted in Figure~\ref{fig:jhks}
are red crosses, which show the synthetic colours of the stellar sequence derived
from the Bruzual-Persson-Gunn-Stryker (BPGS) 
spectroscopic atlas \citep{1983ApJS...52..121G} in Section 1. The synthetic locus is in overall
good agreement with the locus of bright stars in this colour-colour space.

\begin{figure}
\begin{center}
\advance\leftskip-0.8cm
\includegraphics[scale=0.4]{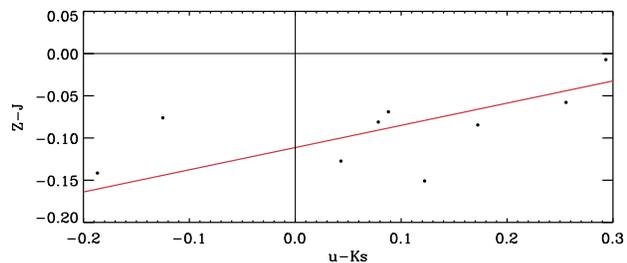}
\caption{Maximum likelihood fit to stars with Vega like colours 
  in SDSS with counterparts in VIKING. Similar fits were made in
  all other VIKING colour combinations vs. $u$-${\rm Ks}$, the results are summarised
  in Table~\ref{tab:offsets} and in the main text.}
\label{fig:zjuks}
\end{center}
\end{figure}

Following Hodgkin et al., we now extend the baselines
on each ${\rm Z}$, ${\rm Y}$, ${\rm J}$, ${\rm H}$ and ${\rm Ks}$ colour-colour combination in one direction by plotting all 
possibilities against $u - \mathrm{Ks}$. Offsets from zero are then measured from
straight line fits to the data. To clarify this procedure an example of the fitting
in the ${\rm Z}$, ${\rm J}$, $u$ and ${\rm Ks}$ colour plane is shown in Figure~\ref{fig:zjuks}. Due to limited statistics our approach to
the fitting differs slightly from \citet{2009MNRAS.394..675H} who bin their 
data and take measurement errors based on standard deviations on binned median values.
Conversely our approach is to fit to the photometrically selected A0 stars and 
retrieve errors directly from the maximum likelihood fitting procedure. The results
show appreciable offsets in the ${\rm Z}$ and ${\rm Y}$ bands only. These are summarised in 
Table~\ref{tab:offsets}. The corresponding offsets found in the ${\rm J}$, ${\rm H}$ and ${\rm Ks}$ bands are consistent with zero;
$\Delta_{\mathrm{JH}} = 0.006\, \pm 0.004$, $\Delta_{\mathrm{JKs}} = 0.008\, \pm 0.005$, 
$\Delta_{\mathrm{HKs}} = 0.013\, \pm 0.006$. 

\begin{figure}
\advance\leftskip 0.4cm
\includegraphics[scale=0.4]{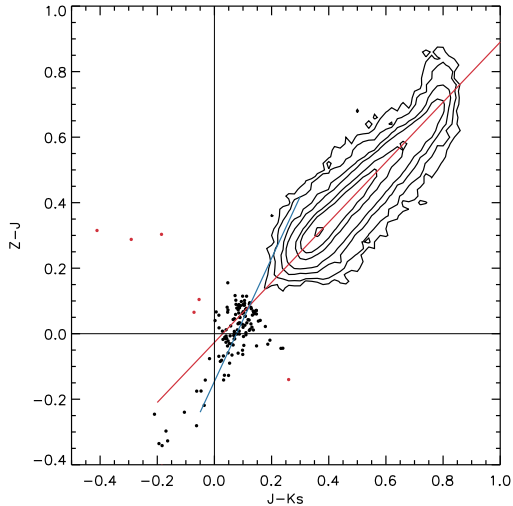}
\caption{The VIKING ${\rm Z}$, ${\rm J}$, ${\rm Ks}$ colour plane for high confidence stellar sources matched in VIKING
  and SDSS. A sample of blue stars is plotted with filled circles, obvious outliers were
  clipped from the sample and are shown in red. The blue line shows
  the maximum likelihood straight line fit to this distribution, while the red line shows
  the fit to the main sample. The ${\rm Z}$ band offset is measured from the blue star intercept
  to the origin.}
\label{fig:zjks}
\end{figure}

A further two results are shown in Table~\ref{tab:offsets}, these were obtained by 
fitting to the blue end of the stellar locus from all the bright
stars in the original selection. Again a straight line maximum likelihood fit is
applied in the ${\rm Z}$, ${\rm J}$, ${\rm Ks}$ and ${\rm Y}$, ${\rm J}$, ${\rm Ks}$ colour-colour 
spaces and the offsets from the origins
are measured. The best fitting line to the ${\rm Z}$, ${\rm J}$, ${\rm Ks}$ stellar locus is shown in Figure~\ref{fig:zjks} and 
the measured offsets in both the ${\rm Z}$, ${\rm J}$, ${\rm Ks}$ and ${\rm Y}$, ${\rm J}$, ${\rm Ks}$ colour spaces 
are summarised in Table~\ref{tab:offsets}. In these cases the measured offsets are slightly larger
than those measured in the VISTA-SDSS colour-colour spaces. These may well be attributed
to small offsets in the ${\rm J}$ or ${\rm Ks}$ passbands as tentatively suggested by 
Figure~\ref{fig:jhks}. Individual offsets to the ${\rm J}$, ${\rm H}$ and ${\rm Ks}$ passbands implied by these fits are
clearly smaller than those attributed to ${\rm Z}$ and ${\rm Y}$ and will not present significant problems in 
the following sections. 
 
\begin{table}
\begin{center}
\begin{tabular}{ccc}
\hline
                                               &   $\Delta_\mathrm{Z}$    &   $\Delta_\mathrm{Y}$  \\ \hline
$\Delta_\mathrm{ZJ}$,   $\Delta_\mathrm{YJ}$   &     $-0.111\, \pm 0.002$         &           $-0.082\,  \pm 0.002$ \\
$\Delta_\mathrm{ZH}$,   $\Delta_\mathrm{YH}$   &     $-0.109\, \pm 0.004$         &           $-0.086\,  \pm 0.004$ \\
$\Delta_\mathrm{ZKs}$,  $\Delta_\mathrm{YKs}$  &     $-0.096\, \pm 0.006$         &           $-0.073\,  \pm 0.005$ \\ 
$\Delta_\mathrm{ZJKs}$, $\Delta_\mathrm{YJKs}$ &     $-0.146\, \pm 0.003$         &           $-0.135\,  \pm 0.003$ \\  \hline
Mean                                           &     $-0.116\, \pm 0.008$         &           $-0.094\,  \pm 0.007$ \\  \hline

\end{tabular}
\end{center}
\caption{Summary of ${\rm Z}$ and ${\rm Y}$ offsets from the Vega system.}
\label{tab:offsets}
\end{table}

\section{Modelling the VIKING Survey}\label{sec:model}

Having used synthetic photometry to predict the colours of cool stars 
and quasars in the VISTA ${\rm Z}$, ${\rm Y}$ and ${\rm J}$ bandpasses, the next step is to combine
this information with number count models of the cool-star and quasar 
populations to produce a simulation of the ${\rm Z}$, ${\rm Y}$ and ${\rm J}$ band photometry expected from
VIKING. 

\subsection{Quasar number counts}

It is well established that the space density of quasars peaks at 
\mbox{${\rm z} \simeq 2.5$} and declines rapidly thereafter 
\citep[e.g.~][]{2006AJ....131.2766R}. By $\mathrm{z} \sim 6$ there
are $\sim 400$ quasars brighter than $z = 21.0$ over the entire sky. 
Since the aim of this work is to limit our follow-up observations to a manageable number of
objects,
the specific number of quasars expected from VIKING is not directly relevant
because the overwhelming majority of follow-up candidates will not be quasars at all.
Nevertheless, the calculation is important for two other reasons; firstly to demonstrate 
that follow-up at this stage presents a statistically viable prospect of discovering
a quasar and secondly because doing so will allow qualitative constraints to be placed 
on the luminosity function when the VIKING quasar search begins to bear fruit.

The number of quasars brighter than an apparent 
magnitude $m_p$ in some fixed observing passband $p$ over a given redshift 
interval may be calculated via integration of the quasar luminosity function (QLF) 
over comoving volume to a depth given by the absolute magnitude limit of the survey.
The latest determination of the QLF at $\rm{z} \simeq 6$ was undertaken by
\citet{2010AJ....139..906W} and combines discoveries from the CFHQS with the more luminous SDSS main
and deep samples. The input catalogue comprises of 40 quasars sampling the redshift range $5.74 <
\rm{z} < 6.42$. The bright end of the binned luminosity function is well constrained to a power law
with some evidence for a flattening in the slope provided by a single quasar at lower
luminosities. In keeping with work at lower redshifts, where the QLF is well constrained over a
large magnitude interval \citep{2000MNRAS.317.1014B,2009MNRAS.399.1755C,2011ApJ...728L..26G}, the
Willott et al.\ parametric QLF has a double power law form.

At  \mbox{$\rm{z} \simeq 6$}, the space density of bright SDSS quasars ($M_{1450} \lesssim -27$) measured by \citet{2001AJ....122.2833F} is consistent 
with a single power law extrapolation of the QLF between $3 \le \rm{z} \le 5$ \citep{1995AJ....110...68S}. The rate of evolution implied by these luminosity functions corresponds to a value $k = -0.47$, where the space density declines exponentially as $10^{k(\rm{z}-6)}$. Beyond \mbox{$\rm{z} \simeq 6$} the value of $k$ is of course completely unknown and moreover one cannot rule out the possibility of a noticeable luminosity dependence, similar to those inherent in quasar, AGN and galaxy populations at lower redshifts \citep{1996AJ....112..839C,2005AJ....129..578B,2009MNRAS.399.1755C}. Clearly then, by extrapolating much past \mbox{$\rm{z} \simeq 6$} one should only make baseline projections.

\begin{figure}
\advance\leftskip0.4cm
\includegraphics[scale=0.4]{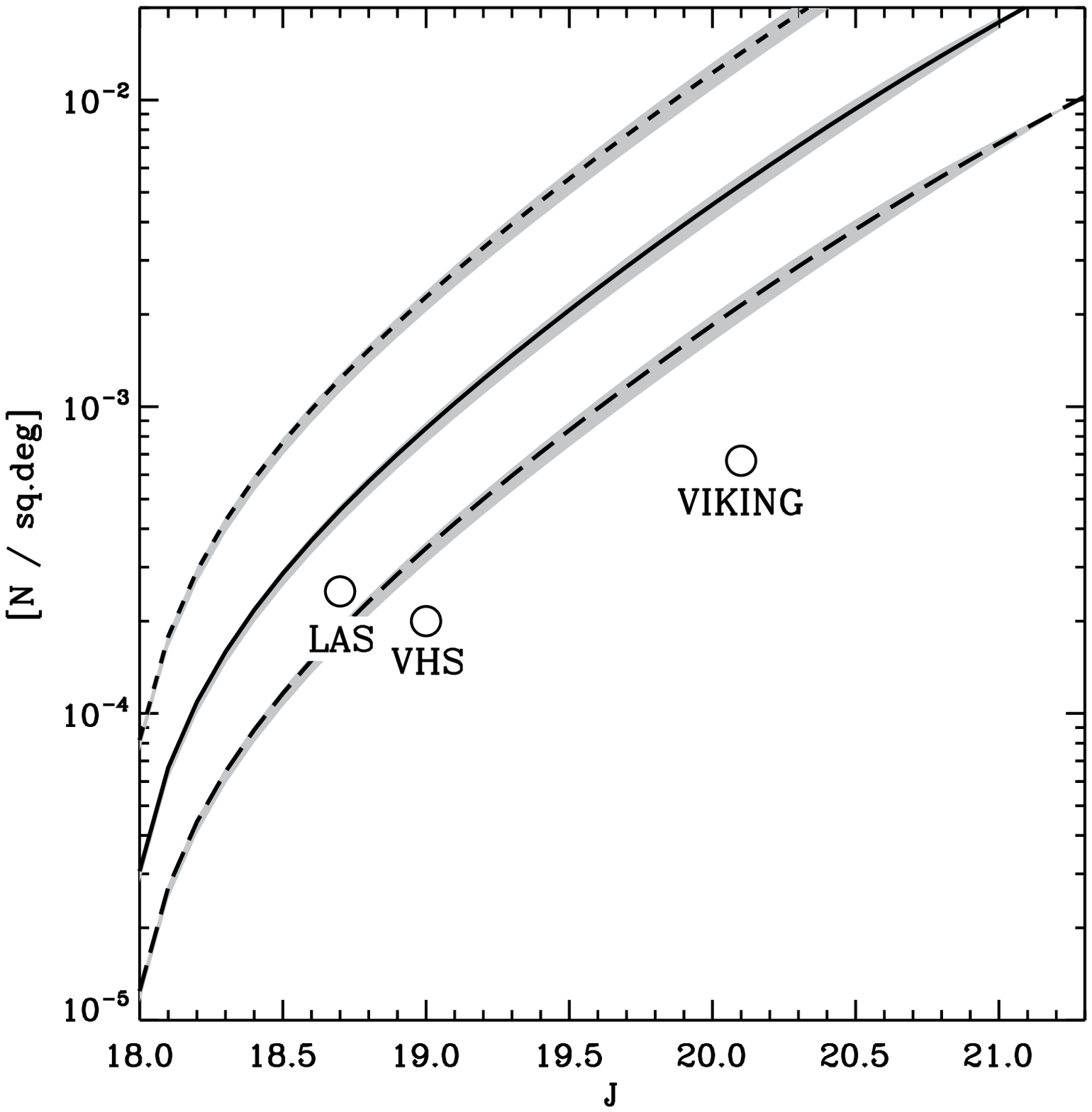}
\caption{The number of \mbox{$6.5 \leq {\rm z} \leq 7.5$} quasars brighter than ${\rm J}$, for three models 
 based on the luminosity function of \citet{2010AJ....139..906W}. The 
 parameter $k$ determines the rate of density evolution beyond 
 \mbox{$\mathrm{z} = 6$}. The short-dashed, solid  and long-dashed curves correspond
 to $k=(0.0,\,-0.47,\,-0.94)$ receptively (see text for details). Open circles have been 
 placed at the inverse area of various surveys, showing the surface 
 density at which the given survey would contain a mean of 1 quasar 
 brighter than its 10$\sigma$ detection limit. Shaded regions
 show the deviations in each curve when the faint end slope ($\beta = -1.5$)
 is allowed to vary between $-2.0 \leq \beta \leq -1.0$.}\label{fig:qcounts}
\end{figure}

We compute a baseline number count model adopting the value $k = -0.47$; we also compute the extreme
cases $k = 0$ (no evolution model) and $k = -0.94$, which corresponds to a factor 3 decline in
density per 0.5 increase in redshift. Model number count curves are shown in
Figure~\ref{fig:qcounts}, which also includes open circles at the inverse area of VIKING, the
UKIDSS-LAS and the VISTA Hemisphere Survey (VHS) at their respective $10\sigma$ sensitivity limits
i.e. the surface density at which the given survey would contain a mean of one quasar. Shaded
regions indicate the change in number counts when the faint end slope of the QLF $(\beta = -1.5)$ is
allowed to vary between $-2.0 \leq \beta \leq -1.0$. The small deviations from the initial models
show that VIKING will probe the bright end of this specific QLF. The baseline model predicts $\sim
8~\mathrm{z} \geq 6.5$ quasars brighter than the $10\sigma$ sensitivity limit over the entire VIKING field, which would correspond to a total of $\sim$2 quasars in the first $350\,\mathrm{deg}^2$ of imaging.

 The \mbox{UKIDSS} eighth data release (\mbox{DR8}) contains some \mbox{ 2600 deg$^2$} of \mbox{LAS}
 imaging in at least the ${\rm Y}$- and ${\rm J}$-bands. The baseline number count model predicts that there should be a mean of two $\mathrm{z} \geq 6.5$ quasars in this area, while the non-evolving model predicts \mbox{$\sim 3$}. At present there is one reported discovery of a $\mathrm{z} \geq 6.5$ quasar (${\rm z} \simeq 7$ to be a little more precise) in \mbox{UKIDSS} \mbox{DR8} \citep{2011Natur.474..616M}, which marginally favours the evolving models.

\subsection{Stellar number counts}

The Besan{\c c}on stellar population synthesis model of the Galaxy 
\cite[BGM;][]{2003A&A...409..523R,2003ASPC..287..104R} allows stellar
number counts up to spectral type M to be computed as function of ${\rm J}$-band magnitude and 
direction on the sky via an interactive web form\footnote{http://model.obs-besancon.fr/}. The model is currently 
undergoing an update to the set of M-dwarf atmospheres it uses. The current study therefore
provides an ideal testbed in which to compare their performance. Here we briefly review the
model inputs and the changes to the M-dwarf atmospheres.

The atmospheres of cooler L- and T-dwarfs are much less well constrained and so we take a
more empirical approach to calculate their number counts.

\subsubsection{The Besan{\c c}on Stellar Population Synthesis Model of the Galaxy: M-dwarf number counts}
The Besan{\c c}on Galaxy Model has been designed to produce simulations of the stellar content in
the Galaxy as seen towards different lines of sight and photometric bands. It is based on assumptions
about stellar evolution, Galactic evolution and Galactic dynamics. The model includes 4 populations
(thin disc, thick disc, halo and bulge) each one deserving a specific treatment. The main
ingredients to simulate those populations are the density laws, the initial mass function (IMF) and
the star formation rate.

At the magnitudes considered here, the more relevant parameters to estimate counts of M-dwarf stars are the slope of the IMF at low masses and the scale height. Thin disc scale height has been obtained from a dynamical argument; the scale height depends on the local density, the age-velocity dispersion relation and the potential through the Boltzmann equation. Hence the thin disc scale height is fixed for an isothermal population of a given age and a given vertical velocity dispersion.
The different scale heights or ellipticities as a function of age are given in \cite{2003A&A...409..523R}.

The IMF used in the simulations is different from the BGM on line, which was obtained in an analysis in
\cite{2006A&A...447..185S} where the slope of the IMF in the thick disc has been found to be very
steep. Later on this IMF was discarded when compared with deep data because it overestimated
numbers of M-dwarfs in the simulations. This mistake was probably due to a default of the calibration
of the model in the Canada France Hawaii Legacy Survey (CFHTLS) bands. The IMF slope has now been 
revised and we use here a slope of $\alpha=0.7$ down to very low masses, as in
\cite{1997A&A...320..440H} 
where the IMF is parametrised as $dN/dM \propto M^{-(1+\alpha)}$. 

The computations of the magnitudes and colours of the stars rely on the stellar atmosphere
models. Despite of the fact that the Basel library was used for most the HR (Hertzsprung-Russell) diagram \citep{1997A&AS..125..229L}, for low mass stars those models significantly deviate from observations \citep{2006A&A...447..185S}. Instead we use NextGen model atmospheres \citep{1997ARA&A..35..137A} at effective temperatures smaller than about $4000\,K$. At the present time, the BGM does not include brown dwarfs of masses smaller than the hydrogen burning limit.

Stellar number counts were calculated on a 2D grid of spectral-type and ${\rm J}$-mag bins for 
30 representative fields of $10\,\mathrm{deg}^2$ for each of the thin-disc, thick-disc and halo components. The 
results were scaled up accordingly. In order to make certain that density gradients were 
well sampled, effort was made to ensure that each field was offset from adjacent fields 
by no more than $5\,\mathrm{deg}$ of Galactic latitude. ${\rm J}$-band number counts were then
calculated in the range $16.0 \le \mathrm{J} \le 24.0$ in bin widths of $\Delta \mathrm{J} = 0.5$. 
According to the quasar number count model this is $\sim\,2$ mag brighter than we can 
realistically expect to recover quasars in VIKING and $\sim\,3$ mag fainter than the 
5$\sigma$ limiting magnitude of the survey. The extra range is included to allow
sources to scatter in and out of the survey as the result of photometric error.

\subsubsection{L- and T-dwarf number counts}

For number counts of L- and T-dwarfs one can consider the standard empirical Galactic
density model, whereby the total contribution from the thin and thick discs is 
given by the sum of two double exponential profiles \cite[e.g.][]{2001ApJ...553..184C}. A full description of
the number count calculation is provided by \cite{2008A&A...488..181C} so here only the main
points are reviewed.

 The space density of either disc component 
 at Galactic coordinates ($l$, $b$) and heliocentric distance 
 $d$ assumes the following form 

\begin{equation}\label{eqn:thindisc}
n(d,l,b)\propto \mathrm{exp}
 \Bigg(\,-\frac{R(d,l,b)-R_{\odot}} {L}
 \Bigg)\, 
 \mathrm{exp}
 \,\Bigg({-\frac{|Z_\odot+d\,\mathrm{sin}\,b|} {{H}}}\Bigg), 
\end{equation} 

\noindent  where $Z_\odot$ is the position of the sun relative 
 to the Galactic plane and  
 ${H}$ and ${L}$ represent the scale height and length
  of the particular disc component. Since VIKING observes extra-galactic fields 
  i.e. out of the Galactic plane, the assumption \mbox{$R_\odot \gg d$} is valid 
  and the galactocentric distance 
  at the point of interest $R(d,l,b)$, is given by the approximation

\begin{equation}
\label{eqn:disttopoint}
R(d,l,b)\approx R_\odot - d\,\mathrm{cos}\,b\,\mathrm{cos}\,l.
\end{equation}  

\noindent The proportionality constant 
 is the spatial density of the relevant object class or 
 classes at \mbox{$Z=0$} i.e.~at the Galactic plane. 

We adopt the local space density and absolute ${\rm J}$--band magnitude per
spectral type estimates of \citet{2008A&A...488..181C} and calculate stellar number counts per spectral type 
 in ${\rm J}$-magnitude bins of width 0.5 over the range \mbox{$16 \le \rm J \le 24$}.
The parameters for the thin and thick discs are taken from
 \citet{2001ApJ...553..184C} and \citet{2008ApJ...673..864J} as given in Table~\ref{tab:thindisc}.

We assume an identical luminosity function 
 in both the thin and thick discs. 
This probably overestimates the count
 of thick-disk brown dwarfs, since old population sub-stellar objects 
 are predicted to have very faint magnitudes by theoretical cooling 
sequences \citep{2005ARA&A..43..195K}. This would place most 
 of the thick-disk population 
 with \mbox{$M < 0.06 M_\odot$} below \mbox{VIKING} detection limits. 
 This assumption is therefore conservative since it will 
  overestimate the brown-dwarf contamination of the quasar sample. 

A similar argument can be made for the halo, which represents an even older population. In
this case we assume that all halo L- and T-dwarfs will be below \mbox{VIKING} detection limits.
BGM number counts of halo M-dwarfs fall off rapidly towards cooler spectral types. The BGM surface 
density of halo M9-dwarfs with ${\rm J} < 23.0$ is $< 1\:{\rm deg^{-2}}$, suggesting that this 
is a reasonable approach.

\begin{table}\caption{Parameters of the Galactic 
thin and thick disc system \citep{2001ApJ...553..184C,2008ApJ...673..864J}}\label{tab:thindisc}
\begin{center}
\begin{tabular}{c|c|l|c|c}
\hline
              $R_\odot$ & $Z_\odot$ & & L    & H \\
              (pc)      & (pc)      & & (pc) &(pc)\\
\hline
             8600 $\pm$ 200 & +27 $\pm$ 4 & Thin: & 2250 $\pm$ 1000  &330 $\pm$ 3 \\
             &              & Thick: & 3600 $\pm$ 720 & 580 - 750 \\              
&& \multicolumn{3}{|c|}{Thick disc normalisation: 13 - 6.5 \%}	      \\
\hline
\end{tabular}
\end{center}
\end{table}

\subsection{Generating Synthetic Observations}

With stellar number counts binned by apparent $\rm J$-band magnitude and spectral
type, the next step is to pair each count with a set of synthetic photometry to
mimic an observation of each source.

For $\rm N$ sources distributed over ${\rm J}_1 \dots {\rm J}_p \dots {\rm J}_P$ magnitude
and ${\rm T}_1\dots {\rm T}_q \dots {\rm T}_Q$ spectral type intervals, the procedure is to step over all 
$P \times Q$ bins iteratively. Each object receives a set of synthetic fluxes 
given by $\mbox{\boldmath $F$}=(f_{\rm Z},\,f_{\rm Y},\,f_{\rm J})$.
$\mbox{\boldmath $\widetilde{F}$}$ is then generated via interpolation 
of our synthetic photometry (Section 2) between a source of 
the relevant spectral type ${\rm T}_q$
and that of the later adjacent spectral type, ${\rm T}_{q+1}$. If the
$r\,{\rm th}$ occupant of bin $p,\, q$ has a magnitude $\rm J$ in the range
${\rm J}_p \leq {\rm J} < {\rm J}_{p+1}$ and a spectral type ${\rm T}_q$, then
$\mbox{\boldmath $\widetilde{F}$}$ is defined by the following weighted sum:

\begin{equation}
\mbox{\boldmath $\widetilde{F}$}= \alpha\,\mbox{\boldmath $F$}_{p,\,q,\,r} + (1-\alpha)\,\mbox{\boldmath $F$}_{p,\,q+1,\,r},
\end{equation}

\noindent where $\alpha$ is a uniform random deviate in the range $0 < \alpha \leq 1$. 
Each of the components of $\mbox{\boldmath$\widetilde{F}$}$ (i.e.\
$\mbox{\boldmath $F$}_{p,\,q,\,r}$ and $\mbox{\boldmath $F$}_{q,\,q+1,\,r}$), comprise
of $\rm Z$-, $\rm Y$- and $\rm J$-band fluxes generated by drawing $\rm Z$- and $\rm Y$-band fluxes from our library of synthetic 
photometry for the relevant spectral type. These are then scaled to a reference flux 
$f_{\rm J}$, which is derived from a magnitude 
$\rm J$ drawn randomly from the interval ${\rm J}_p \leq {\rm J} < {\rm J}_{p+1}$. 
This process is repeated over all $P \times Q$ bins until all number counts are coupled
to a set of unique synthetic photometry.

The VSA VIKING release v1.0 currently contains 173 image tiles. During the image reduction and
pipeline processing the average sky noise in each image tile is measured by placing a number of 
circular apertures in sparsely populated regions. These
measurements are then placed in the header units of each image before being uploaded to the VSA.
In a real survey of faint sources the distribution of measured flux errors is sky dominated and usually
well approximated as being Gaussian. The intrinsic distribution of synthetic fluxes can therefore
be perturbed by drawing from a set of Gaussian distributions, with standard deviations equal to the 
sky noise level in the VIKING image tiles and scaled to the theoretical flux of the
object in the corresponding passband. Adding this distribution of errors to the set of simulated 
intrinsic fluxes gives a set of simulated `observations'.    

\subsection{Generating a Synthetic Catalogue}

The catalogue extraction process for the VIKING survey, is handled by the 
VISTA Data Flow System (VDFS) at the Cambridge Astronomical Survey Unit (CASU). 
The CASU source extraction algorithm does not implement a rigid detection threshold.
Thus to determine whether our synthetic
sources would be detected by the CASU source extractor it is necessary to model the
S/N distribution of detected sources from the data. Here we follow
a novel method used by \citet{PatelTheses}, which is described in brief below.

Since we are modelling a population of point sources it is necessary to limit the data
analysis to stars only, which requires a highly reliable method of star-galaxy separation.
In contrast to \citet{PatelTheses}, who used morphological classifiers to define a reliable
and complete sample of point-like objects, here we use colour separation by matching sources 
in VIKING, the VISTA Deep Extra-galactic Observations survey (VIDEO) and the CFHTLS. 
These surveys share an overlap region of $\sim 1\, \mathrm{deg}^2$ 
in the VIDEO tile centred on $\mathrm{Ra} = \mathrm{02^h\,26^m\,00^s}$ and $\mathrm{dec} =
\mathrm{-04^\circ\,30^{\prime}\,00^{\prime \prime}}$
(the VIDEO-XMM3 field). 
VIDEO observes in the same passbands as VIKING 
but is $\sim 2.5$ mags deeper ($5\sigma$ limiting depths of ${\rm Z}=25.2$, ${\rm Y}=24.0$, ${\rm
  J}=23.6$, ${\rm H}=22.6$ and ${\rm Ks}=21.7$).
At the time of writing only the ${\rm Y}$, ${\rm J}$, ${\rm H}$ and ${\rm Ks}$ observations are available, but this is not important
since most high-z quasars will be too faint to be detected in VIKING ${\rm Z}$-band observations given
the CASU $\sim 5\sigma$ extraction criterion. 

\begin{figure}
\begin{center}
\advance\leftskip-0.8cm
\includegraphics[scale=0.4]{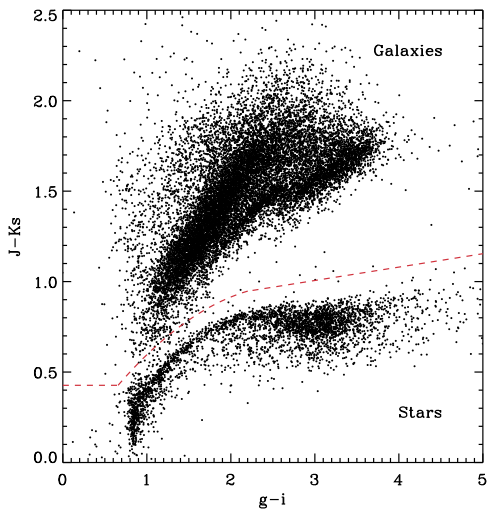}
\caption{The CFHTLS-VIDEO $g$, $i$, ${\rm J}$, ${\rm Ks}$ colour plane for matched VIKING-VIDEO-CFHTLS sources
  in the $\sim 1\,\mathrm{deg}^2$ region of overlap shared by each survey. The deep
  CFHTLS-VIDEO observations allow robust star galaxy separation. The red dashed line shows
  the regions in which the star and galaxy populations are defined.
}\label{fig:gijks}
\end{center}
\end{figure}

The CFHTLS Deep survey observes in the Sloan passbands to limiting depths of
$u\!=\!\mathrm{26.6},~g\!=\!\mathrm{28.0},~r\!=\!\mathrm{27.6},~i\!=\!\mathrm{27.0}~\mathrm{and}~z\!=\!\mathrm{25.7}$.
Sources matched between all three surveys are plotted in the CFHTLS-VIDEO 
$g - i$ vs. $\mathrm{J} - \mathrm{Ks}$ plane in Figure~\ref{fig:gijks}. The extra depth in the matched catalogue
minimises photometric scatter and, with the combination of colours, gives remarkably
good star galaxy separation. The red dashed line in Figure~\ref{fig:gijks} separates two regions from which we draw a sample
of stars and galaxies. Where possible $i$ or $g$ dropouts have been classified by virtue of either 
their $z-\mathrm{J}$ vs $\mathrm{J-Ks}$ colours or their $\mathrm{J-Ks}$ colours alone. In this
section we concentrate on the stellar sample.

\begin{figure}
\begin{center}
\advance\leftskip-0.8cm
\includegraphics[scale=0.4]{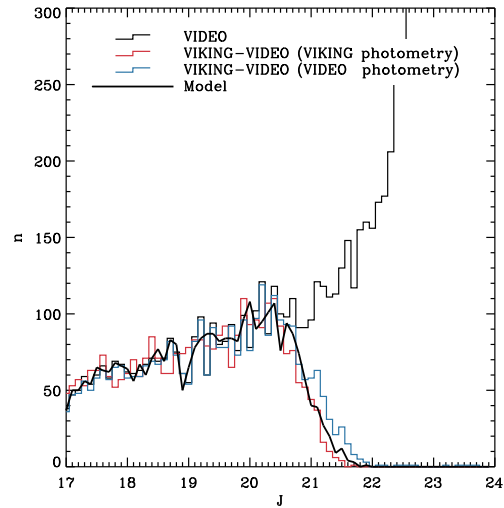}
\caption{The ${\rm J}$ band magnitude distributions of sources matched between VIDEO-CFHTLS (black histogram) and
  VIKING-VIDEO-CFHTLS with VIKING photometry (red histogram) and with VIDEO photometry (blue histogram).
  The black continuous curve shows the distribution of simulated sources with intrinsic magnitudes
  drawn from the VIDEO-CFHTLS matched catalogue and perturbed with realistic VIKING-like Gaussian errors.
  A Gaussian distributed signal to noise cut has been applied to mimic the source 
  extraction process used to produce the VIKING catalogue (see text for details).
}\label{fig:multiphotJ}
\end{center}
\end{figure}

Histograms of the ${\rm J}$-band magnitude distributions of matched stars are shown in Figure~\ref{fig:multiphotJ}
for VIKING (red) and VIDEO (blue) photometry.
For comparison, the full distribution of VIDEO-CFHTLS matched point sources is also plotted (black histogram).
Synthetic sources are drawn randomly from this distribution and each source is 
perturbed by a Gaussian error with a standard deviation equal to the sky-noise level drawn from 
VIKING image tiles. Each perturbed source is then given a detection significance (S/N) threshold drawn 
from a Gaussian distribution with mean $\mu_p$ and standard deviation $\sigma_p$. 
If a given source possesses a S/N greater than it simulated detection threshold then it is kept otherwise 
it is discarded. The resulting distribution is then compared by eye to the VIKING-VIDEO-CFHTLS matched 
distribution and the process is repeated with various values of $\mu_p$ and $\sigma_p$
until a good match is found. The continuous black curve in Figure~\ref{fig:multiphotJ} shows the accepted matched
${\rm J}$-band distribution
with the parameters $(\mu_\mathrm{J}$, $\sigma_\mathrm{J})=(0.22,\:0.04)$. 
The same procedure was undertaken in the ${\rm Y}$-band yielding a Gaussian with parameters
 $(\mu_\mathrm{Y}$, $\sigma_\mathrm{Y})=(0.15,\:0.06)$.
The findings were applied to the set of simulated observations described above
resulting in a realisation of the cool star component of the
VIKING catalogue over the entire $1500\,\mathrm{deg}^2$ of the VIKING field. 

\section{The Z, Y, J Selection Criteria}\label{sec:selection}

\begin{figure*}
\advance\leftskip-0.8cm
\begin{tabular}{cc}
\includegraphics[scale=0.63]{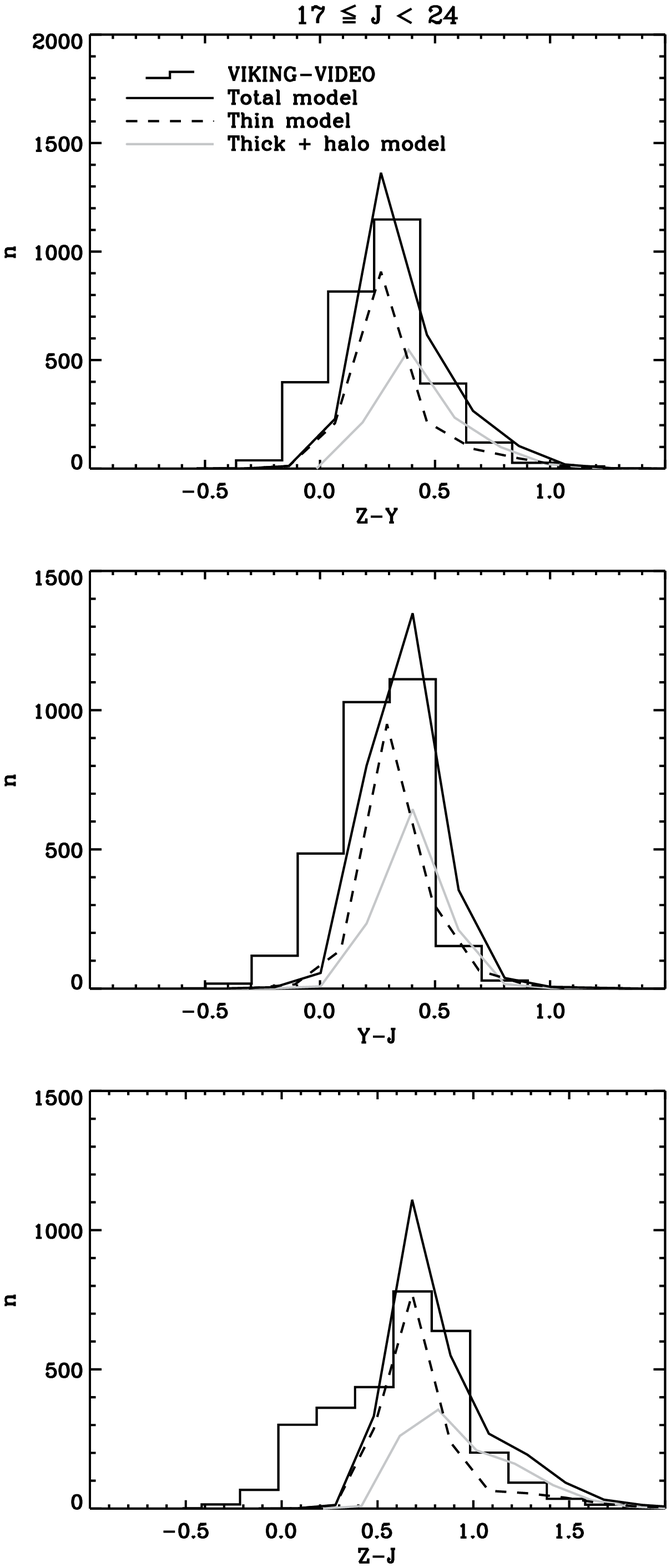}&
\includegraphics[scale=0.63]{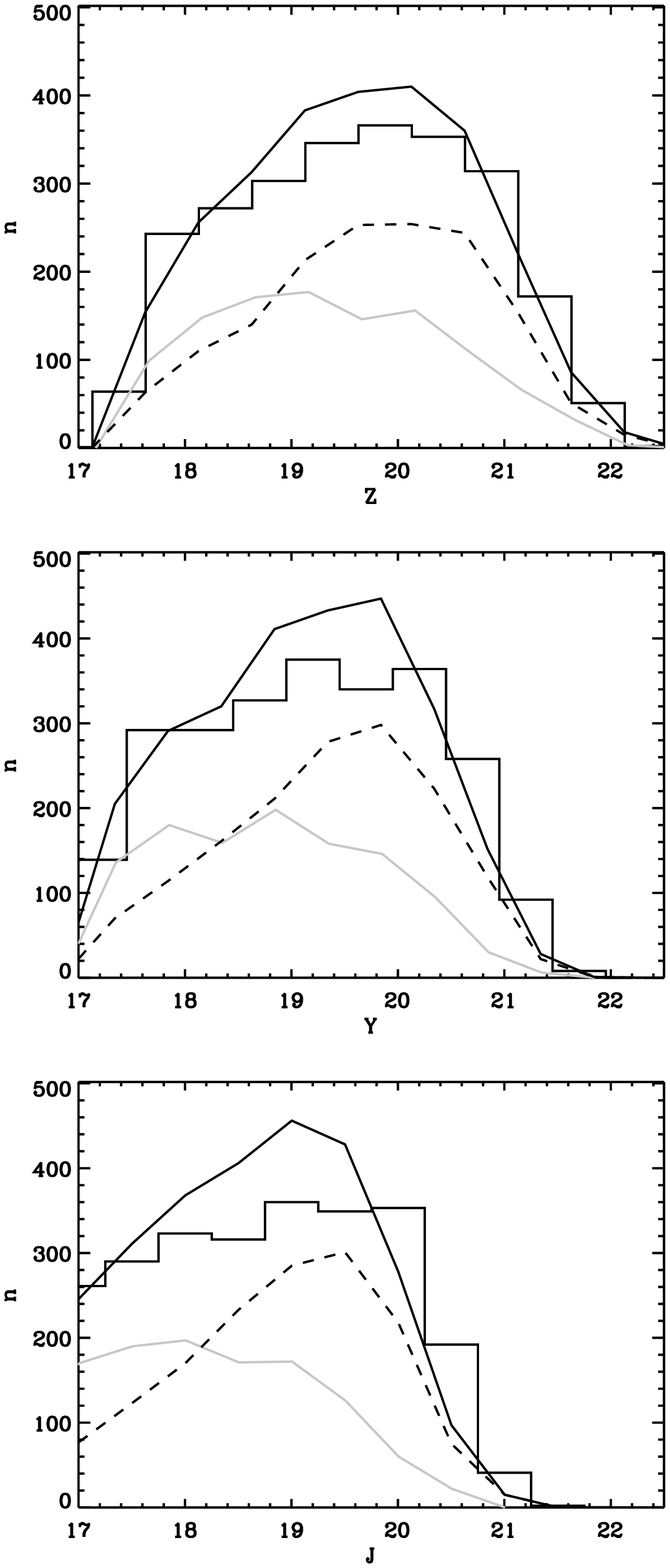}
\end{tabular}\caption{Left: VIKING ${\rm Z}$, ${\rm Y}$, ${\rm J}$ colour distributions of the simulated thin disc, 
  combined thick disc and halo and total cool star contributions in $\sim 1\,\mathrm{deg}^2$ of 
  the VIKING-VIDEO-CFHTLS overlap region. Each distribution is compared to the 
  real distribution of VIKING stellar photometry as shown by the histogram
  style plots. The excess of blue objects in each plot is attributed to main-sequence halo
  stars with bright absolute magnitudes, an effort has been made to remove these objects in order to
  compare modelled and true magnitude distributions. Right: As in the left hand panel but for the simulated and real 
  ${\rm Z}$, ${\rm Y}$, ${\rm J}$ magnitude distributions. In both cases the the overall agreement with the data
  is good. There is a slight overshoot in predicted number counts making these models conservative.}\label{fig:dist}
\end{figure*}

The magnitude and colour distribution of the simulated stellar population has been 
compared with data from the VIKING-VIDEO-CFHTLS catalogue. The results are shown
in Figure~\ref{fig:dist}, where we plot the real data in histogram format and the simulated
catalogue as smooth curves. An excess of blue stars can be seen in the colour distributions in the 
left hand panels. Closer inspection of the magnitude distribution of these stars shows that they
are reasonably faint with number counts peaking at ${\rm Z} \sim 19.5$, 
${\rm Y} \sim 19.5$ and ${\rm J} \sim 19.0$. Given their blue colours, they are likely  
intrinsically bright main sequence halo stars. To promote a fair comparison between the true and 
modelled magnitude distributions an effort has been made to remove the majority of the blue 
stars from the data by retaining only those having 
\mbox{${\rm Z}-{\rm Y} \ge 0.0$, ${\rm Y}-{\rm J} \ge 0.0$, ${\rm Z}-{\rm  J} \ge 0.3$}.

The overall agreement between colour and magnitude distributions in the 
model and the data is good, but the model over predicts 
the number counts in some bins by up to a factor of 1.3.
The number count predictions are therefore conservative.
In the modelled population almost all objects are M-dwarfs whose number counts and 
${\rm J}$-band magnitudes are predicted by the BGM. Given the rarity of cooler type objects, 
the $1\,{\rm deg^2}$ patch of deep VIKING-VIDEO-CFHTLS matched imaging is not sufficient 
to constrain the L- and T-dwarf number count model well. The overall good agreement with 
the data suggests that for the latest type dwarfs, the modelled objects are distributed
accurately over colour and magnitude space and that the predicted number counts are within
an order of magnitude of their true values. It is also encouraging that our 
results broadly conform with data from recent near-IR searches 
\citep{2010MNRAS.406.1885B,2009MNRAS.397..258L}. With a set of well placed colour cuts,
L- and T-dwarfs are known to make only modest contributions to high-z quasar candidate 
lists in comparison to M-dwarfs, so the propagation of these uncertainties will not 
have a large impact on estimates of quasar colour space contamination.

In the following section the modelled stellar and quasar photometry are employed
along with data from the VIKING-VIDEO-CFHTLS overlap region to investigate 
contamination to quasar colour space and to lay down some benchmark selection 
criteria relevant to the first candidate follow-up run. We emphasise the use of
the word `benchmark' here to mean that we will be working conservatively, placing
tighter constraints than will perhaps be necessary. In a real survey it is often useful
to relax constraints e.g. so that one can consider the errors on candidates
around the selection cutoff boundaries.

\subsection{Contamination of quasar colour space}

Take as an illustrative example the first follow-up of VIKING quasar candidates, which was scheduled over three nights
in June 2011. As a rough estimate one might expect this to be sufficient to make 
$\sim 100$ observations. Prior to this, there was expected to be $\sim \mathrm{300} \,\mathrm{deg}^2$ 
of data available from which to choose candidates. This allowed us to set a first 
broad-brush selection constraint e.g. that we must limit our follow-up catalogue to 
$\sim \mathrm{1}$ candidate per 3 $\mathrm{deg}^2$. To place further, similar benchmark constraints 
it is necessary to look in more detail at contamination of quasar colour space.

\subsubsection{Extra-galactic Contamination}

Up to now, the focus of this work has been on contamination from foreground Galactic
M-, L- and T-dwarfs. Referring back to Figure~\ref{fig:zyj} it is clear that contamination
from Galaxies could also pose a problem. Fortunately most galaxies can be rejected
on the basis of their extended disc-bulge morphologies. At faint magnitudes, it is
possible for the galactic disc component to fall below the detection limit of the survey,
while the bulge component remains bright. Galaxies detected in this way can be morphologically
indistinguishable from stars and usually the best way to avoid this type of contamination
is to apply a detection significance constraint to ensure that candidates are sufficiently
well detected in at least one band.

The star galaxy separation techniques discussed in the previous section present the opportunity 
to study the statistical properties of VIKING galaxies in detail and place further selection
constraints on both the morphology and detection significance of quasar colour space contaminants. 
The resulting galaxy sample is on the whole slightly less reliable than the stellar sample, since 
it shares a similar region of colour space with low-z ($0.1 \leq \mathrm{z} \leq 2.5$) quasars. 
The lack of any SDSS spectroscopic overlap prevents any attempts to remove these interlopers
with any confidence, but work on optical and near-IR number counts \citep[e.g.][]{2006AJ....131.2766R,2008MNRAS.386.1605M} 
suggests that low-z quasars will be outnumbered by galaxies in this sample by at least two orders of 
magnitude and they can be reasonably neglected.

The VSA provides a number of morphological classifiers based on the curve of growth of
the individual detections of each source in a set of progressively inclusive apertures. 
The finer details of this approach are discussed extensively by Irwin~et~al.\ (In preparation), the main
points are summarised as follows; measurements are made on ellipticity and curve-of-growth
statistics. The results are then compared at the detector level to the average stellar locus as a function
of magnitude. In this way each detection is given a morphological classification statistic known in 
the VSA as the mergedClassStat (MCS). For stars, this statistic is well approximated by an $N(0,1)$ 
Gaussian distribution. Each individual classification is combined to give the overall classification
for each source.

\begin{figure}
\advance\leftskip-0.2cm
\includegraphics[scale=0.47]{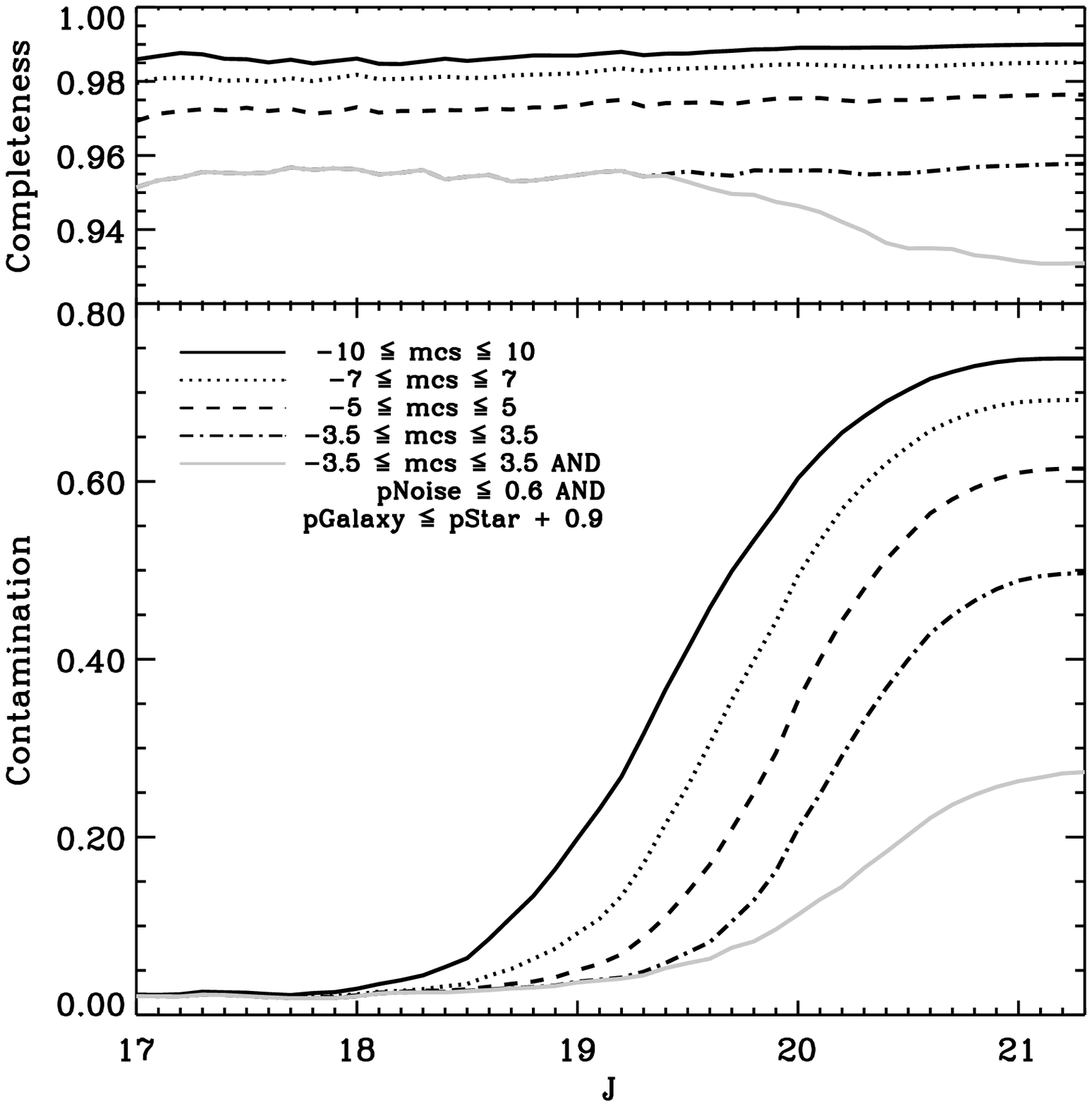} \\
\caption{Upper: The completeness of VIKING-VIDEO-CFHTLS stellar sample with various
  morphological selection constraints imposed (see legend). Lower: The contamination of the
  VIKING-VIDEO-CFHTLS stellar sample by galaxies and noise with various morphological selection 
  constraints imposed. All morphological statistics are derived from VIKING imaging.}\label{fig:cc}
\end{figure}

At bright magnitudes stars and galaxies populate two separate and well defined MCS loci. At
faint magnitudes the galactic locus becomes indistinguishable from the stellar locus. We add to the 
sample of stars and galaxies, a sample of objects which are unmatched between VIKING and VIDEO and
label these sources as `noise'. Figure~\ref{fig:cc}
shows the completeness and contamination of the stellar sample by galaxies and noise within the following constraints; 
$-10 \leq \mathrm{MCS} \leq 10$ (solid curve), $-7 \leq \mathrm{MCS} \leq 7$ (dotted curve),
$-5 \leq \mathrm{MCS} \leq 5$ (dashed curve) and $-3.5 \leq \mathrm{MCS} \leq 3.5$ (dash-dotted curve).
It is clear that losses in completeness
are small compared to gains in minimising contamination. However even with the tightest
constraints contamination contributes to $\sim 60$ per cent of the sample
at the faintest magnitudes.   

The VSA supplies a second useful morphological classification statistic called the
pClass. pClass is derived from a discrete statistic which allocates 
a classification based on the passband level MCS and
several catches to make classification more reliable (e.g.~for saturated sources).
Each discrete classification corresponds to a reasonably accurate, self-consistent 
probability for each of the allowed cases; star, galaxy, noise and saturated.
For each source the pClass statistics for each separately available detection
are then combined using Bayesian classification rules assuming independence. 
This defines the probability  that a given source is either a star, galaxy, noise or 
saturated, the relevant attributes in the VSA being pStar, pGalaxy, pNoise and pSaturated.

The sources in our matched VIKING-VIDEO-CFHTLS catalogue are all fainter than 
$\mathrm{J} = 17.0$ and so in all cases $\mathrm{pSaturated} = 0.0$ and 
$\mathrm{pStar} + \mathrm{pGalaxy} + \mathrm{pNoise} = 1.0$. 

We place the following cuts on each of the morphological classifiers,

\begin{eqnarray}\label{eqn:morphcut}
\lefteqn{\mathrm{pNoise} \leq 0.6 \nonumber}\\
\lefteqn{\mathrm{pGalaxy} \leq \mathrm{pStar} + 0.9 }\nonumber
\end{eqnarray}

\noindent and in Figure~\ref{fig:cc} the contamination and completeness of the stellar sample is
plotted in grey for $-3.5 \leq \mathrm{MCS} \leq 3.5$. With the new
constraints the contamination from galaxies is significantly reduced with only
a small reduction in completeness as the penalty.

\begin{figure}
\includegraphics[scale=0.5]{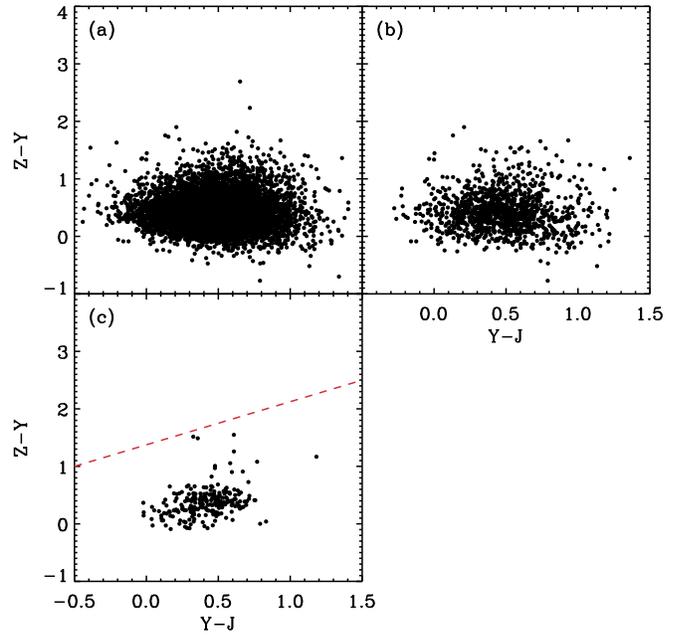}
\caption{The VIKING ${\rm Z}$, ${\rm Y}$, ${\rm J}$ colour plane for a highly complete sample of galaxies
  selected from the VIKING-VIDEO-CFHTLS overlap region. Panel (a) shows all galaxies,
  Panel (b) shows remaining galaxies after morphological cuts have been imposed and
  Panel (c) shows the sample after both morphological cuts and an 8$\sigma$ detection
  significance threshold in ${\rm Y}$ and ${\rm J}$ is applied. The red dashed line shows a conservative
  colour selection constraint, which rejects all remaining galaxies.}\label{fig:gzyj}
\end{figure}

Figure~\ref{fig:gzyj}~(a) plots all galaxies in the ${\rm Z}$, ${\rm Y}$, ${\rm J}$ plane and
galaxies passing the morphological constraints are plotted in panel (b). 
Given the morphological constraints the remaining galaxies in (b) are all
indistinguishable from stars.
We naturally want to select quasars as faint as possible, but  
experience from other surveys has shown that an \mbox{$8\sigma$} to 
\mbox{$10\sigma$} detection in at least one  band is necessary to avoid the selection of large numbers 
of false positives \citep[e.g.][]{2001AJ....122.2833F}. Therefore at the faint end we adopt a nominal photometric cut at 
$\mathrm{S/N} \geq 8$ in both ${\rm Y}$ and ${\rm J}$. Galaxies in Figure~\ref{fig:gzyj} passing this constraint are plotted in
panel (c). Only a small fraction of the original galaxies remain and most are concentrated within the main galaxy locus. 
There are perhaps half a dozen sources lying sufficiently far from the galaxy locus to be 
mistaken for potential candidates. Since this is now the realm of small number statistics
it is impossible to draw strong conclusions on the frequency of these or similar objects.
Assuming that this distribution is typical for all VIKING galaxies, then in the worst case scenario
all outliers from the locus would be considered for followup. This suggests, perhaps
$\sim 10$ extra-galactic contaminants per ${\rm deg}^2$, although this is clearly a weak
constraint. Better constraints can be placed on the extent of extra-galactic contamination
when we begin to identify our candidates spectroscopically. For now, it is safer to work
conservatively by placing a set of benchmark colour cuts such that all galaxies are rejected. 
These cuts are shown by the red dashed line in
Figure~\ref{fig:gzyj}. 

\subsubsection{Stellar Contamination}

\begin{figure}
\advance\leftskip-0.4cm
\includegraphics[scale=0.55]{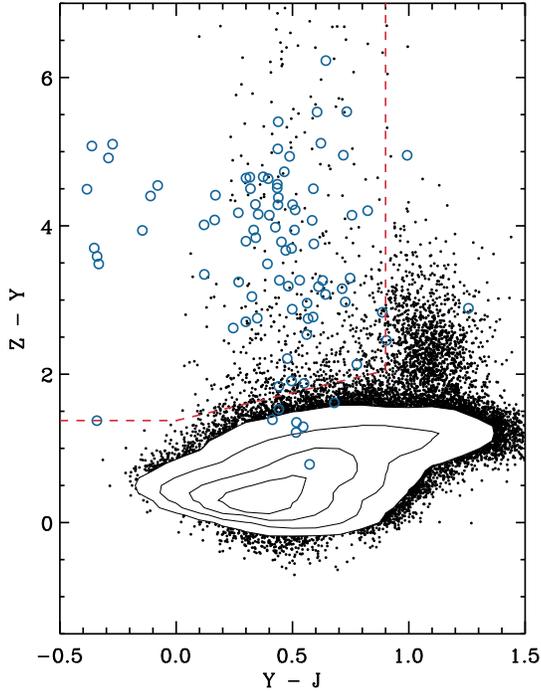}
\caption{The VIKING ${\rm Z}$, ${\rm Y}$, ${\rm J}$ colour plane for simulated cool stars (contours and points) in the VIKING catalogue.
  Also shown are $\sim 200$ simulated $6.5 \leq {\rm z} \leq 7.5$ quasars (open circles) drawn from a uniform
  distribution in redshift and `detected' at the $8\sigma$ level in ${\rm Y}$ and ${\rm J}$. The red
  dashed line defines a benchmark colour selection 
  region that encloses $\sim 1$ object per $3~\mathrm{deg}^2$ as was required by the conditions of our
  follow-up campaign.}\label{fig:qzyj}
\end{figure}

Figure~\ref{fig:qzyj} plots the simulated cool-star catalogue in ${\rm Z}$, ${\rm Y}$, ${\rm J}$ colour-colour space.
Stars are represented as black points and contours.
Plotted in blue as open circles is a catalogue of $\sim 200$ simulated $6.5 \leq {\rm z} \leq 7.5$ 
quasars. Each source has been
`detected' at the $8\,\sigma$ level in Y and J and was drawn from larger catalogue of
$\sim 1000$ quasars with a uniform distribution in redshift. Given the
selection space constraints derived in the previous section, the task here is simple; 
to refine these criteria to allow
$\sim 1$ interloper per $3\,\mathrm{deg}^2$ while simultaneously ensuring that
the vast majority of quasars lie within these constraints. One also must ensure that the
revised constraints do not constitute a relaxation of the above constraints.

The revised selection region is described by Equation~\ref{eqn:zyjcut} and shown in Figure~\ref{fig:qzyj} by the
dashed line, which encloses $\sim 500$ interloping stars.
As required, this corresponds to $\sim 1$ interloper per $3\,\mathrm{deg}^2$ over the 
entire $1500\,\mathrm{deg}^2$ of the VIKING field. 

\begin{eqnarray}
  {\rm Z}-{\rm Y} &\geq& 1.375 \nonumber
  \\ {\rm Y}-{\rm J} &\leq& 0.900 \label{eqn:zyjcut} 
  \\ {\rm Z}-{\rm Y} &\geq& 0.750\,({\rm Y}-{\rm J})+1.375 \nonumber
\end{eqnarray}

Finally it is extremely important to verify whether these cuts 
maintain high completeness.

\subsection{Selection Completeness}\label{sec:complete}

\begin{figure}
\advance\leftskip-0.5cm
\includegraphics[scale=0.5]{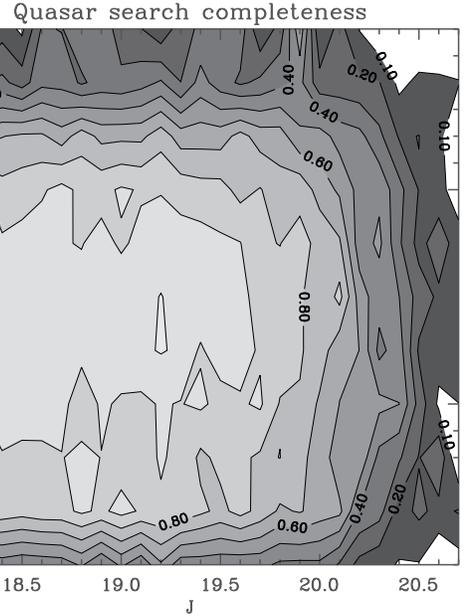}
\caption{Estimated VIKING survey completeness for the benchmark selection constraints. Contours are
  plotted in ${\rm J}$ vs. z and increase in intervals on 0.1. The selection function accounts
  for loss of completeness from all significance, colour and morphological constraints
  described in the text. }\label{fig:qcomp}
\end{figure}

The estimated completeness contours of the selection strategy
are shown as a function of ${\rm J}$ and redshift in Figure~\ref{fig:qcomp}.
Completeness is defined here as the fraction of catalogued quasars of
intrinsic ${\rm J}$-band magnitude $\mathrm{J}$ and redshift $\mathrm{z}$,
with observables that pass the morphological, significance and colour
constraints outlined above. The total mean completeness of the survey is
$\sim\,60$ per cent and is robust against moderate changes in the selection criteria
that maintain the $\sim 1$ interloper per $3\,\mathrm{deg}^2$ requirement.

The conservative selection constraints imposed on the simulated catalogue maintain a
completeness that is comparable to but slightly lower than the highly successful SDSS 
and CFHTLS quasar searches in the optical and the UKIDSS search in the near-IR
\citep{2003AJ....125.1649F,2008AJ....135.1057J,2010AJ....139..906W,PatelTheses}. The
difference can be accounted for firstly by our conservative selection constraints and
secondly because completeness estimations in these searches do not consider the losses
due to morphological rejection. That said,
it is important to mention that the completeness will be further affected by a few 
additional factors not modelled here. Firstly, loss
 of completeness due to survey `holes' caused by ghosts 
 around bright stars; 
 secondly, blending with foreground objects causing genuine quasars
 to move bluewards in \mbox{${\rm Z-Y}$} and be rejected  
 and finally, non-Gaussian errors and spurious detections 
 causing increased contamination compared to our models. 

The bright-star losses are expected to be relatively small, rejecting
 typically a few percent of the sky.  Secondly, the losses
 from chance projection of foreground objects are also modest;
 at \mbox{$\rm Z \sim 23$} the surface density of objects is \mbox{$\sim 10 $
  per arcmin$^2$}, so if we assume (conservatively) real quasars will be rejected
  if closer than 2 arcsec to such an object, this will cause
   an inefficiency of 3.5 per cent. 

Finally, there are many sources of non-Gaussian errors which 
 may create spurious ${\rm Z}$-dropouts, including moving objects such
 as asteroids, image persistence, hot pixels, etc. 
 The \mbox{VIKING} observing strategy ensures that each object is
 observed at approximately 8 jitter positions on different pixels, and also
that the ${\rm J}$-band imaging is split into two epochs separated by 
 weeks or months. Using this, we expect that almost all 
  spurious, time-variable and moving objects can be rejected by checking 
   of individual data frames, though the
  precise number remains to be quantified in detail.

With a mean completeness of $\sim 60$ per cent, the quasar number count
models derived in Section~\ref{sec:model}, suggest that $\sim 2$ $6.5 \leq {\rm z} \leq 7.5$
quasars brighter than the VIKING $8\sigma$ point-source sensitivity limit will be selected from 
the first $\sim 350 {\rm\, deg}^2$ of VIKING imaging, by the benchmark selection constraints.

\section{NTT Imaging Observations}\label{sec:ntt}

The first VIKING quasar candidates were observed in ${\rm I_{NTT}}$ and
${\rm Z_{NTT}}$ imaging by the ESO Faint Object Spectrograph and Camera II
(EFOSC2) on ESO's New Technology Telescope\footnote{Based on observations made 
with ESO Telescopes at the La Silla Paranal Observatory under programme ID 087.A-0655(A)}
(NTT) over a three night 
period between 26th - 29th June 2011. In practise both traditional 
colour selection and probabilistic candidate ranking \citep{2011arXiv1101.4965M} 
were employed independently to converge on the final candidate list and as recommended 
in Section~\ref{sec:selection} the benchmark colour and significance constraints were relaxed considerably in order 
to remain inclusive. Here we examine the VIKING ${\rm Z}$, ${\rm Y}$, ${\rm J}$ photometry of these candidates and discuss
implications for the benchmark selection constraints in light of the NTT 
observations. 

In total 129 science exposures were taken of 44 candidates\footnote{
Given the modest area covered by VIKING at this 
stage ($\sim350\,{\rm deg}^2$), it should be noted that the observation 
run was heavily dictated by the visibility of the patchy VIKING footprint
throughout the night. Thus the observations presented here do not 
represent an exhaustive list of our highest priority candidates.}. 
All candidates were observed in ${\rm I_{NTT}}$ to depths of $\gtrsim 22.8$ 
mag at the $3 \sigma$ noise level. At these depths ${\rm z} \geq 6.5$ quasars 
are dark in optical passbands so candidates showing a visible flux peak in these 
images were discarded. In total just 6 candidates appeared undetected 
in ${\rm I_{NTT}}$ images that reached $3 \sigma$ magnitude limits of 
${\rm I_{NTT}} > 23.7$ (this has since been confirmed with aperture photometry).

\begin{figure}
\advance\leftskip-0.5cm
\includegraphics[scale=0.5]{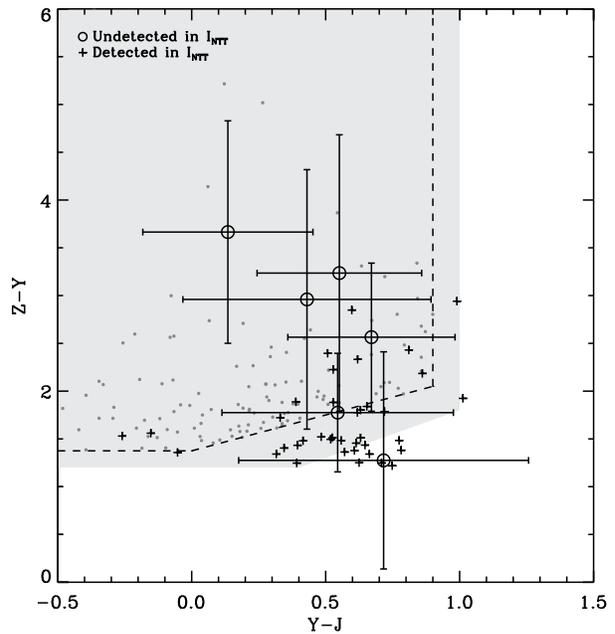}
\caption{High-z quasar candidates in the VIKING ${\rm Z}$, ${\rm Y}$, ${\rm J}$ colour plane. The dashed line
shows the benchmark colour selection region. Grey circles show all VIKING candidates
falling within this region after applying the benchmark morphological and photometric 
constraints described in the main text. A fraction of these objects are high priority 
candidates awaiting follow-up but most were rejected as result of probabilistic
ranking or due to optical associations with SDSS sources. Crosses and open circles are
candidates observed in ${\rm I_{NTT}}$ and ${\rm Z_{NTT}}$ . Open circles were undetected in 
${\rm I_{NTT}}$ and have ${\rm I_{NTT}}$, ${\rm Z_{NTT}}$, ${\rm Y}$, ${\rm J}$ colours consistent
with $6.5 \leq {\rm z} \leq 7.0$ 
quasars. Error bars show $1\sigma$ uncertainties.}\label{fig:zyjfol}
\end{figure}

In Figure~\ref{fig:zyjfol} the dashed line shows the benchmark selection region 
in the VIKING ${\rm Z}$, ${\rm Y}$, ${\rm J}$ colour-colour plane. The filled grey region shows the set of relaxed
constraints used in the search. 
Dark grey points represent objects which pass the benchmark
constraints but have not undergone follow-up. A small fraction remain strong 
candidates and are awaiting observation, but most were rejected after 
probabilistic ranking or because they are associated with an optical source in the 
SDSS overlap region. Crosses and open circles show ${\rm I_{NTT}}$ visible and 
${\rm I_{NTT}}$ drop-out candidates respectively. As is clear from the scatter of
these points outside of the grey region, even the relaxed constraints 
were not strictly adhered to.

The ${\rm I_{NTT}}$ drop-out lying at ${\rm Z-Y} \simeq 1.25, {\rm Y-J} \simeq 0.7$
is a faint (${\rm S/N} \simeq 6$ in ${\rm Y}$) low-priority candidate and would not normally 
have been observed but for a nightly period in which most high priority candidates 
were not visible. Given its low priority it was not observed in ${\rm Z_{NTT}}$. Excluding this
object there remain five strong ${\rm I_{NTT}}$ drop-out candidates which were subsequently 
followed up in ${\rm Z_{NTT}}$ in order to confirm a break shortward of the VISTA ${\rm Y}$-band. 
All were found to have a combination of colours in ${\rm I_{NTT}}$,
 ${\rm Z_{NTT}}$, ${\rm Y}$ and ${\rm J}$ consistent with $6.5 \leq {\rm z} \leq 7.0$ quasars. 
 Four of these candidates lie firmly within the
benchmark colour selection constraints and a fifth lies on the boundary,
three candidates lie just outside the significance constraints
(S/N $>$ 7.3; which serves as a reminder of the importance of widening
the search space in order to consider the colour uncertainties on objects
lying around the selection boundaries).
The clustering of these candidates in and around the selection 
region provides considerable support  for the choice of selection constraints.
A detailed discussion and analysis of the NTT photometry, as well as the spectroscopy
of these candidates is given by Venemans et al.~(2011).

\section{Summary and Future Work}\label{sec:sum}

The first public data release of the VIKING survey will be delivered to ESO in mid-
to late-2011.
Approximately $ \mathrm{75 \, deg}^2$ is expected to be made public.
 VIKING is expected to contain a significant sample
of $6.5 \leq \mathrm{z} \leq 7.5$ quasars, the precise number depending 
sensitively on the unknown evolution of the quasar luminosity function 
beyond \mbox{${\rm z} \sim 6$}. 

Follow-up of the first VIKING quasar candidates was conducted in June 2011.
Candidates were chosen from $\sim 350\,{\rm deg}^2$ of data available at the 
time. To conduct a full and rigorous search of this data set a selection strategy 
was required in advance of the availability of most science ready data products.
This made it difficult to develop a robust set of quasar 
selection constraints which optimise both efficiency in photometric and spectroscopic 
follow-up, whilst simultaneously selecting a highly complete sample of quasars.

To remedy this, we have modelled the cool star component of the VIKING catalogue in the 
${\rm Z}$, ${\rm Y}$, ${\rm J}$ passbands to mimic contamination of quasar colour space by dwarf stars
of spectral type M, L and T. We have compared this model to 
$\sim 1\,\mathrm{deg}^{2}$ of imaging in the VIKING-VIDEO-CFHTLS overlap region
using photometry from the complementary $g,\:i,\:\mathrm{J,\:Ks}$ colours as
a robust star-galaxy separator. The model and data are in good agreement and
we use it along with a sample of galaxies from the same overlapping data set 
 to place constraints on the photometric and morphological characteristics
of quasar colour space contaminants. This allows us to define a set of benchmark
photometric and morphological selection cuts which will limit follow-up candidates
to $\sim 1$ object per $3\,\mathrm{deg}^2$, an initial requirement of the follow-up
campaign. The completeness of this search strategy is comparable to 
similar quasar searches conducted both in the optical and the near-IR.

Follow-up imaging of the first quasar candidates from the NTT provides considerable support for the choice 
of benchmark constraints. Of 44 candidates observed in either ${\rm I_{NTT}}$ or
both ${\rm I_{NTT}}$ and ${\rm Z_{NTT}}$, just 5 have colours consistent with 
$6.5 \leq {\rm z} \leq 7.5$ quasars. All 5 of these sources have VIKING colours
that place them within our benchmark selection constraints.

Several other optical and near-IR surveys are due to either see their first light or
to release their first products in the near future (e.g. Pan-STARRS, VST-KIDS, VHS). 
The tendency with large surveys is to place priority initially on the timely release 
of a small volume of data from a few well studied fields (VIKING being no exception). 
The methodology described here will provide an equally complementary set of methods with 
which to assess the first data from these forthcoming catalogues, allowing benchmark
constraints to be placed on all rare object searches. 

As surveys and corresponding data volumes begin to enter the Peta-byte era, rare object searches 
employing simple colour cuts are becoming
cumbersome and inefficient. Several authors have already begun to prioritise follow-up
candidates by combining observables to place them in a well defined ranking 
scheme \citep{2009AJ....137.3884R,2010A&A...523A..14Y,2011ApJ...729..141B}. The first such ranking 
algorithm for high-z quasar candidates, uses Bayesian
model comparison techniques to determine the probability that a given source has scattered
from the quasar locus \citep{2011arXiv1101.4965M}. In principle this methodology does away with colour cuts all together
retaining and treating information that is normally lost or considered qualitatively, in a
robust mathematical framework. The algorithm presented by Mortlock et al. requires accurate
models of the photometry of cool stars and quasars specific to the particular survey it is
applied to. Ideally the stellar model is constrained via maximum likelihood fits to the
real data. In the case of VIKING and several other surveys this is not
possible with the limited data expected from initial releases. 

In a forthcoming paper we will
present quasar candidates selected from the VSA by applying the stellar model developed here to 
the Mortlock et al. algorithm. We will discuss the success of this approach to modelling the stellar population
by characterising a sample of test set data. We will also discuss in detail, the techniques necessary
for defining the input catalogue from VSA archived photometry. Our hope is that this will provide a starting point
from which to encourage future independent quasar searches in the VIKING data.  
This shows the diversity of the tools developed here in
the context of high-z quasar and other rare object searches.  

\section{ACKNOWLEDGEMENTS}

The authors would like to thank Jim Emerson and Karim Malik 
for their support and helpful discussion. We would also like to thank the UK team responsible for the realisation of VISTA, and the ESO team who operate and maintain this new facility. The UK's VISTA Data Flow System comprising the VISTA pipeline at the Cambridge Astronomy Survey Unit  (CASU) and the VISTA Science Archive at the Wide Field Astronomy Unit (Edinburgh) (WFAU) provided calibrated data products and is supported by STFC.

The Besan\c{c}on Simulations have been executed on computers from the Utinam Institute
of the Universit\'e de Franche-Comt\'e, supported by the R\'egion de
Franche-Comt\'e and Institut des Sciences de l'Univers (INSU)

We also make use of data based on observations obtained with MegaPrime/MegaCam, a joint project of CFHT and CEA/DAPNIA, at the Canada-France-Hawaii Telescope (CFHT) which is operated by the National Research Council (NRC) of Canada, the Institut National des Science de l'Univers of the Centre National de la Recherche Scientifique (CNRS) of France, and the University of Hawaii. This work is based in part on data products produced at TERAPIX and the Canadian Astronomy Data Centre as part of the Canada-France-Hawaii Telescope Legacy Survey, a collaborative project of NRC and CNRS. 

\fontsize{8pt}{10pt}
\selectfont


\begin{thebibliography}{}

\bibitem[\protect\citeauthoryear{{Allard}, {Hauschildt}, {Alexander} \&
  {Starrfield}}{{Allard} et~al.}{1997}]{1997ARA&A..35..137A}
{Allard} F.,  {Hauschildt} P.~H.,  {Alexander} D.~R.,    {Starrfield} S.,
  1997, \araa, 35, 137

\bibitem[\protect\citeauthoryear{{Barger}, {Cowie}, {Mushotzky}, {Yang},
  {Wang}, {Steffen} \& {Capak}}{{Barger} et~al.}{2005}]{2005AJ....129..578B}
{Barger} A.~J.,  {Cowie} L.~L.,  {Mushotzky} R.~F.,  {Yang} Y.,  {Wang} W.,
  {Steffen} A.~T.,    {Capak} P.,  2005, \aj, 129, 578

\bibitem[\protect\citeauthoryear{{Bochanski}, {Hawley}, {Covey}, {West},
  {Reid}, {Golimowski} \& {Ivezi{\'c}}}{{Bochanski}
  et~al.}{2010}]{2010AJ....139.2679B}
{Bochanski} J.~J.,  {Hawley} S.~L.,  {Covey} K.~R.,  {West} A.~A.,  {Reid}
  I.~N.,  {Golimowski} D.~A.,    {Ivezi{\'c}} {\v Z}.,  2010, \aj, 139, 2679

\bibitem[\protect\citeauthoryear{{Bovy}, {Hennawi}, {Hogg}, {Myers},
  {Kirkpatrick}, {Schlegel}, {Ross}, {Sheldon}}{{Bovy} et~al.}{2011}]{2011ApJ...729..141B}
{Bovy} J.,  {Hennawi} J.~F.,  {Hogg} D.~W., et al., 2011, \apj, 729, 141

\bibitem[\protect\citeauthoryear{{Boyle}, {Shanks}, {Croom}, {Smith}, {Miller},
  {Loaring} \& {Heymans}}{{Boyle} et~al.}{2000}]{2000MNRAS.317.1014B}
{Boyle} B.~J.,  {Shanks} T.,  {Croom} S.~M.,  {Smith} R.~J.,  {Miller} L.,
  {Loaring} N.,    {Heymans} C.,  2000, \mnras, 317, 1014

\bibitem[\protect\citeauthoryear{{Burningham}, {Pinfield}, {Lucas}, {Leggett},
  {Deacon}, {Tamura}, {Tinney}, {Lodieu} et~al.,}{{Burningham}
  et~al.}{2010}]{2010MNRAS.406.1885B}
{Burningham} B.,  {Pinfield} D.~J.,  {Lucas} P.~W., et~al., 2010, \mnras,
  406, 1885

\bibitem[\protect\citeauthoryear{{Caballero}, {Burgasser} \&
  {Klement}}{{Caballero} et~al.}{2008}]{2008A&A...488..181C}
{Caballero} J.~A.,  {Burgasser} A.~J.,    {Klement} R.,  2008, \aap, 488, 181

\bibitem[\protect\citeauthoryear{{Chen}, {Stoughton}, {Smith} et~al.,}{{Chen}
  et~al.}{2001}]{2001ApJ...553..184C}
{Chen} B.,  {Stoughton} C.,  {Smith} J.~A.,    et~al., 2001, \apj, 553, 184

\bibitem[\protect\citeauthoryear{{Cowie}, {Songaila}, {Hu} \& {Cohen}}{{Cowie}
  et~al.}{1996}]{1996AJ....112..839C}
{Cowie} L.~L.,  {Songaila} A.,  {Hu} E.~M.,    {Cohen} J.~G.,  1996, \aj, 112,
  839

\bibitem[\protect\citeauthoryear{{Croom}, {Richards}, {Shanks}, {Boyle},
  {Strauss}, {Myers}, {Nichol}, {Pimbblet}, et~al.}{{Croom} et~al.}{2009}]{2009MNRAS.399.1755C}
{Croom} S.~M.,  {Richards} G.~T.,  {Shanks} T., et al.,  2009, \mnras, 399, 1755

\bibitem[\protect\citeauthoryear{{Emerson}, {Irwin}, {Lewis}, {Hodgkin},
  {Evans}, {Bunclark}, {McMahon}, {Hambly} et~al.,}{{Emerson}
  et~al.}{2004}]{2004SPIE.5493..401E}
{Emerson} J.~P.,  {Irwin} M.~J.,  {Lewis} J., et~al., 2004, in
  {P.~J.~Quinn \& A.~Bridger} ed., Society of Photo-Optical Instrumentation
  Engineers (SPIE) Conference Series Vol.~5493 of Society of Photo-Optical
  Instrumentation Engineers (SPIE) Conference Series, {VISTA data flow system:
  overview}.
pp 401--410

\bibitem[\protect\citeauthoryear{{Emerson} \& {Sutherland}}{{Emerson} \&
  {Sutherland}}{2010}]{2010Msngr.139....2E}
{Emerson} J.,  {Sutherland} W.,  2010, The Messenger, 139, 2

\bibitem[\protect\citeauthoryear{{Fan}}{{Fan}}{1999}]{1999AJ....117.2528F}
{Fan} X.,  1999, \aj, 117, 2528

\bibitem[\protect\citeauthoryear{{Fan}, {Narayanan}, {Lupton} et~al.,}{{Fan}
  et~al.}{2001}]{2001AJ....122.2833F}
{Fan} X.,  {Narayanan} V.~K.,  {Lupton} R.~H.,    et~al., 2001, \aj, 122, 2833

\bibitem[\protect\citeauthoryear{{Fan}, {Strauss}, {Schneider} et~al.,}{{Fan}
  et~al.}{2003}]{2003AJ....125.1649F}
{Fan} X.,  {Strauss} M.~A.,  {Schneider} D.~P.,    et~al., 2003, \aj, 125, 1649

\bibitem[\protect\citeauthoryear{{Fan}, {Strauss}, {Becker} et~al.,}{{Fan}
  et~al.}{2006}]{2006AJ....132..117F}
{Fan} X.,  {Strauss} M.~A.,  {Becker} R.~H.,    et~al., 2006, \aj, 132, 117

\bibitem[\protect\citeauthoryear{{Francis}, {Hewett}, {Foltz}, {Chaffee},
  {Weymann} \& {Morris}}{{Francis} et~al.}{1991}]{1991ApJ...373..465F}
{Francis} P.~J.,  {Hewett} P.~C.,  {Foltz} C.~B.,  {Chaffee} F.~H.,  {Weymann}
  R.~J.,    {Morris} S.~L.,  1991, \apj, 373, 465

\bibitem[\protect\citeauthoryear{{Glikman}, {Djorgovski}, {Stern}, {Dey},
  {Jannuzi} \& {Lee}}{{Glikman} et~al.}{2011}]{2011ApJ...728L..26G}
{Glikman} E.,  {Djorgovski} S.~G.,  {Stern} D.,  {Dey} A.,  {Jannuzi} B.~T.,
  {Lee} K.,  2011, \apjl, 728, L26

\bibitem[\protect\citeauthoryear{{Gunn} \& {Stryker}}{{Gunn} \&
  {Stryker}}{1983}]{1983ApJS...52..121G}
{Gunn} J.~E.,  {Stryker} L.~L.,  1983, \apjs, 52, 121

\bibitem[\protect\citeauthoryear{{Hambly}, {Mann}, {Bond}, {Sutorius}, {Read},
  {Williams}, {Lawrence} \& {Emerson}}{{Hambly}
  et~al.}{2004}]{2004SPIE.5493..423H}
{Hambly} N.~C.,  {Mann} R.~G.,  {Bond} I.,  {Sutorius} E.,  {Read} M.,
  {Williams} P.,  {Lawrence} A.,    {Emerson} J.~P.,  2004, in {P.~J.~Quinn \&
  A.~Bridger} ed., Society of Photo-Optical Instrumentation Engineers (SPIE)
  Conference Series Vol.~5493 of Society of Photo-Optical Instrumentation
  Engineers (SPIE) Conference Series, {VISTA data flow system survey access and
  curation: the WFCAM science archive}.
pp 423--431

\bibitem[\protect\citeauthoryear{{Haywood}, {Robin} \& {Creze}}{{Haywood}
  et~al.}{1997}]{1997A&A...320..440H}
{Haywood} M.,  {Robin} A.~C.,    {Creze} M.,  1997, \aap, 320, 440

\bibitem[\protect\citeauthoryear{{Hewett}, {Foltz} \& {Chaffee}}{{Hewett}
  et~al.}{1995}]{1995AJ....109.1498H}
{Hewett} P.~C.,  {Foltz} C.~B.,    {Chaffee} F.~H.,  1995, \aj, 109, 1498

\bibitem[\protect\citeauthoryear{{Hewett}, {Warren}, {Leggett} \&
  {Hodgkin}}{{Hewett} et~al.}{2006}]{2006MNRAS.367..454H}
{Hewett} P.~C.,  {Warren} S.~J.,  {Leggett} S.~K.,    {Hodgkin} S.~T.,  2006,
  \mnras, 367, 454

\bibitem[\protect\citeauthoryear{{Hodgkin}, {Irwin}, {Hewett} \&
  {Warren}}{{Hodgkin} et~al.}{2009}]{2009MNRAS.394..675H}
{Hodgkin} S.~T.,  {Irwin} M.~J.,  {Hewett} P.~C.,    {Warren} S.~J.,  2009,
  \mnras, 394, 675

\bibitem[\protect\citeauthoryear{{Irwin}, {Lewis}, {Hodgkin}, {Bunclark},
  {Evans}, {McMahon}, {Emerson}, {Stewart} et~al.,}{{Irwin}
  et~al.}{2004}]{2004SPIE.5493..411I}
{Irwin} M.~J.,  {Lewis} J.,  {Hodgkin} S., et~al., 2004, in
  {P.~J.~Quinn \& A.~Bridger} ed., Society of Photo-Optical Instrumentation
  Engineers (SPIE) Conference Series Vol.~5493 of Society of Photo-Optical
  Instrumentation Engineers (SPIE) Conference Series, {VISTA data flow system:
  pipeline processing for WFCAM and VISTA}.
pp 411--422

\bibitem[\protect\citeauthoryear{{Jiang}, {Fan}, {Vestergaard}, {Kurk},
  {Walter}, {Kelly} \& {Strauss}}{{Jiang} et~al.}{2007}]{2007AJ....134.1150J}
{Jiang} L.,  {Fan} X.,  {Vestergaard} M.,  {Kurk} J.~D.,  {Walter} F.,  {Kelly}
  B.~C.,    {Strauss} M.~A.,  2007, \aj, 134, 1150

\bibitem[\protect\citeauthoryear{{Jiang}, {Fan}, {Annis} et~al.,}{{Jiang}
  et~al.}{2008}]{2008AJ....135.1057J}
{Jiang} L.,  {Fan} X.,  {Annis} J.,    et~al., 2008, \aj, 135, 1057

\bibitem[\protect\citeauthoryear{{Jiang}, {Fan}, {Brandt}, {Carilli}, {Egami},
  {Hines}, {Kurk}, {Richards}, {Shen}, {Strauss}, {Vestergaard} \&
  {Walter}}{{Jiang} et~al.}{2010}]{2010Natur.464..380J}
{Jiang} L.,  {Fan} X.,  {Brandt} W.~N.,  {Carilli} C.~L.,  {Egami} E.,  {Hines}
  D.~C.,  {Kurk} J.~D.,  {Richards} G.~T.,  et al.,  2010, \nat, 464, 380

\bibitem[\protect\citeauthoryear{{Juarez}, {Maiolino}, {Mujica}, {Pedani},
  {Marinoni}, {Nagao}, {Marconi} \& {Oliva}}{{Juarez}
  et~al.}{2009}]{2009A&A...494L..25J}
{Juarez} Y.,  {Maiolino} R.,  {Mujica} R.,  {Pedani} M.,  {Marinoni} S.,
  {Nagao} T.,  {Marconi} A.,    {Oliva} E.,  2009, \aap, 494, L25

\bibitem[\protect\citeauthoryear{{Juri{\'c}}, {Ivezi{\'c}}, {Brooks}
  et~al.,}{{Juri{\'c}} et~al.}{2008}]{2008ApJ...673..864J}
{Juri{\'c}} M.,  {Ivezi{\'c}} {\v Z}.,  {Brooks} A.,    et~al., 2008, \apj,
  673, 864

\bibitem[\protect\citeauthoryear{Kaiser, Aussel \& Burke}{Kaiser
  et~al.}{2002}]{2002SPIE.4836..154K}
Kaiser N., et al., 2002, in {J.~A.~Tyson \& S.~Wolff} eds.,
  Society of Photo-Optical Instrumentation Engineers (SPIE) Conference Series
  Vol.~4836 of Society of Photo-Optical Instrumentation Engineers (SPIE)
  Conference Series, {Pan-STARRS: A Large Synoptic Survey Telescope Array}.
p.~154

\bibitem[\protect\citeauthoryear{{Kirkpatrick}}{{Kirkpatrick}}{2005}]{2005ARA&%
A..43..195K}
{Kirkpatrick} J.~D.,  2005, \araa, 43, 195

\bibitem[\protect\citeauthoryear{{Komatsu}, {Dunkley}, {Nolta}
  et~al.,}{{Komatsu} et~al.}{2009}]{2009ApJS..180..330K}
{Komatsu} E.,  {Dunkley} J.,  {Nolta} M.~R.,    et~al., 2009, \apjs, 180, 330

\bibitem[\protect\citeauthoryear{{Lawrence}, {Warren}, {Almaini}, {Edge},
  {Hambly}, {Jameson}, {Lucas} \& {Casali}}{{Lawrence}
  et~al.}{2007}]{2007MNRAS.379.1599L}
{Lawrence} A.,  et~al.,  2007, \mnras, 379, 1599

\bibitem[\protect\citeauthoryear{{Leggett}, {Burningham}, {Saumon}
  et~al.,}{{Leggett} et~al.}{2010}]{2010ApJ...710.1627L}
{Leggett} S.~K.,  {Burningham} B.,  {Saumon} D.,    et~al., 2010, \apj, 710,
  1627

\bibitem[\protect\citeauthoryear{{Lejeune}, {Cuisinier} \& {Buser}}{{Lejeune}
  et~al.}{1997}]{1997A&AS..125..229L}
{Lejeune} T.,  {Cuisinier} F.,    {Buser} R.,  1997, \aaps, 125, 229

\bibitem[\protect\citeauthoryear{{Lodieu}, {Burningham}, {Hambly} \&
  {Pinfield}}{{Lodieu} et~al.}{2009}]{2009MNRAS.397..258L}
{Lodieu} N.,  {Burningham} B.,  {Hambly} N.~C.,    {Pinfield} D.~J.,  2009,
  \mnras, 397, 258

\bibitem[\protect\citeauthoryear{{Lupton}, {Gunn} \& {Szalay}}{{Lupton}
  et~al.}{1999}]{1999AJ....118.1406L}
{Lupton} R.~H.,  {Gunn} J.~E.,    {Szalay} A.~S.,  1999, \aj, 118, 1406

\bibitem[\protect\citeauthoryear{{Maddox}, {Hewett}, {Warren} \&
  {Croom}}{{Maddox} et~al.}{2008}]{2008MNRAS.386.1605M}
{Maddox} N.,  {Hewett} P.~C.,  {Warren} S.~J.,    {Croom} S.~M.,  2008, \mnras,
  386, 1605

\bibitem[\protect\citeauthoryear{{Maiolino}}{{Maiolino}}{2009}]{2009AIPC.1111.%
.160M}
{Maiolino} R., 2009, in Giobbi~G., Tornambe~A., Raimondo~G., Limongi~M., Antonelli~L.~A., Menci~N. \&
Brocato~E., eds,
Vol.~1111 of American Institute of Physics Conference
  Series, {Early metal enrichment in high-redshift quasars}. AIP, Melville, New York.
p.~160

\bibitem[\protect\citeauthoryear{{Mortlock}, {Patel}, {Warren}
  et~al.,}{{Mortlock} et~al.}{2009}]{2009A&A...505...97M}
{Mortlock} D.~J.,  {Patel} M.,  {Warren} S.~J.,    et~al., 2009, \aap, 505, 97

\bibitem[\protect\citeauthoryear{{Mortlock}, {Patel}, {Warren}, {Hewett},
  {Venemans}, {McMahon} \& {Simpson}}{{Mortlock}
  et~al.}{2011}]{2011arXiv1101.4965M}
{Mortlock} D.~J.,  {Patel} M.,  {Warren} S.~J.,  {Hewett} P.~C.,  {Venemans}
  B.~P.,  {McMahon} R.~G.,    {Simpson} C.~J.,  2011, ArXiv e-prints

\bibitem[\protect\citeauthoryear{{Mortlock}, {Warren}, {Venemans}, {Patel},
  {Hewett}, {McMahon}, {Simpson}, {Theuns}, et~al.}{Mortlock et~al.}{2011}]{2011Natur.474..616M}
{Mortlock} D.~J.,  {Warren} S.~J.,  {Venemans} B.~P., et~al.
2011, \nat, 474, 616

\bibitem[\protect\citeauthoryear{Patel}{Patel}{2010}]{PatelTheses}
Patel M.,  2010, PhD thesis, Imperial College London

\bibitem[\protect\citeauthoryear{{Richards}, {Fan}, {Newberg}
  et~al.,}{{Richards} et~al.}{2002}]{2002AJ....123.2945R}
{Richards} G.~T.,  {Fan} X.,  {Newberg} H.~J.,    et~al., 2002, \aj, 123, 2945

\bibitem[\protect\citeauthoryear{{Richards}, {Strauss}, {Fan}
  et~al.,}{{Richards} et~al.}{2006}]{2006AJ....131.2766R}
{Richards} G.~T.,  {Strauss} M.~A.,  {Fan} X.,    et~al., 2006, \aj, 131, 2766

\bibitem[\protect\citeauthoryear{{Richards}, {Deo}, {Lacy}, {Myers}, {Nichol},
  {Zakamska}, {Brunner}, {Brandt}, {Gray}, {Parejko}, {Ptak}, {Schneider},
  {Storrie-Lombardi} \& {Szalay}}{{Richards}
  et~al.}{2009}]{2009AJ....137.3884R}
{Richards} G.~T.,  {Deo} R.~P.,  {Lacy} M., et al.,  2009, \aj, 137, 3884

\bibitem[\protect\citeauthoryear{{Robin} \& {Reyl{\'e}}}{{Robin} \&
  {Reyl{\'e}}}{2003}]{2003ASPC..287..104R}
{Robin} A.~C.,  {Reyl{\'e}} C.,  2003, in {J.~M.~De Buizer \& N.~S.~van der
  Bliek} ed., Galactic Star Formation Across the Stellar Mass Spectrum Vol.~287
  of Astronomical Society of the Pacific Conference Series, {The Initial Mass
  Function at Low Masses}.
pp 104--109

\bibitem[\protect\citeauthoryear{{Robin}, {Reyl{\'e}}, {Derri{\`e}re} \&
  {Picaud}}{{Robin} et~al.}{2003}]{2003A&A...409..523R}
{Robin} A.~C.,  {Reyl{\'e}} C.,  {Derri{\`e}re} S.,    {Picaud} S.,  2003,
  \aap, 409, 523

\bibitem[\protect\citeauthoryear{Rowan-Robinson, Babbedge, {Oliver}
  et~al.,}{Rowan-Robinson et~al.}{2008}]{2008MNRAS.386..697R}
Rowan-Robinson M.,  Babbedge T.,  {Oliver} S.,    et~al., 2008, \mnras, 386,
  697

\bibitem[\protect\citeauthoryear{{Schmidt}, {Schneider} \& {Gunn}}{{Schmidt}
  et~al.}{1995}]{1995AJ....110...68S}
{Schmidt} M.,  {Schneider} D.~P.,    {Gunn} J.~E.,  1995, \aj, 110, 68

\bibitem[\protect\citeauthoryear{{Schneider}, {Hall}, {Richards}
  et~al.,}{{Schneider} et~al.}{2007}]{2007AJ....134..102S}
{Schneider} D.~P.,  {Hall} P.~B.,  {Richards} G.~T.,    et~al., 2007, \aj, 134,
  102

\bibitem[\protect\citeauthoryear{{Schultheis}, {Robin}, {Reyl{\'e}},
  {McCracken}, {Bertin}, {Mellier} \& {Le F{\`e}vre}}{{Schultheis}
  et~al.}{2006}]{2006A&A...447..185S}
{Schultheis} M.,  {Robin} A.~C.,  {Reyl{\'e}} C.,  {McCracken} H.~J.,  {Bertin}
  E.,  {Mellier} Y.,    {Le F{\`e}vre} O.,  2006, \aap, 447, 185

\bibitem[\protect\citeauthoryear{{Shen}, {Strauss}, {Oguri} et~al.,}{{Shen}
  et~al.}{2007}]{2007AJ....133.2222S}
{Shen} Y.,  {Strauss} M.~A.,  {Oguri} M.,    et~al., 2007, \aj, 133, 2222

\bibitem[\protect\citeauthoryear{{Skrutskie}, {Cutri}, {Stiening}
  et~al.,}{{Skrutskie} et~al.}{2006}]{2006AJ....131.1163S}
{Skrutskie} M.~F.,  {Cutri} R.~M.,  {Stiening} R.,    et~al., 2006, \aj, 131,
  1163

\bibitem[\protect\citeauthoryear{{Stiavelli}, {Djorgovski}, {Pavlovsky}
  et~al.,}{{Stiavelli} et~al.}{2005}]{2005ApJ...622L...1S}
{Stiavelli} M.,  {Djorgovski} S.~G.,  {Pavlovsky} C.,    et~al., 2005, \apjl,
  622, L1

\bibitem[\protect\citeauthoryear{{Venemans}, {McMahon}, {Warren},
  {Gonzalez-Solares}, {Hewett}, {Mortlock}, {Dye} \& {Sharp}}{{Venemans}
  et~al.}{2007}]{2007MNRAS.376L..76V}
{Venemans} B.~P.,  {McMahon} R.~G.,  {Warren} S.~J.,  {Gonzalez-Solares} E.~A.,
   {Hewett} P.~C.,  {Mortlock} D.~J.,  {Dye} S.,    {Sharp} R.~G.,  2007,
  \mnras, 376, L76

\bibitem[\protect\citeauthoryear{{Warren} \& {Hewett}}{{Warren} \&
  {Hewett}}{2002}]{2002ASPC..283..369W}
{Warren} S.,  {Hewett} P.,  2002, in {Metcalfe} N.,  {Shanks} T.,  eds, A New
  Era in Cosmology Vol.~283 of Astronomical Society of the Pacific Conference
  Series, {WFCAM, UKIDSS, and z = 7 Quasars}.
p.~369

\bibitem[\protect\citeauthoryear{{Wiese}, {Smith} \& {Glennon}}{{Wiese}
  et~al.}{1966}]{1966atp..book.....W}
        Wiese W. L., Smith M. W., Glennon B. M., 1966, NSRDS-NBS 4, Atomic
        Transition Probabilities. Vol. I: Hydrogen through Neon. A Critical
        Data Compilation. US Department of Commerce, National Bureau of
        Standards, Washington, DC

\bibitem[\protect\citeauthoryear{{Willott}, {Percival}, {McLure}, {Crampton},
  {Hutchings}, {Jarvis}, {Sawicki} \& {Simard}}{{Willott}
  et~al.}{2005\natexlab{a}}]{2005ApJ...626..657W}
{Willott} C.~J.,  {Percival} W.~J.,  {McLure} R.~J.,  {Crampton} D.,
  {Hutchings} J.~B.,  {Jarvis} M.~J.,  {Sawicki} M.,    {Simard} L.,  2005,
  \apj, 626, 657

\bibitem[\protect\citeauthoryear{Willott, Delfosse, Forveille, Delorme \&
  Gwyn}{Willott et~al.}{2005\natexlab{b}}]{2005ApJ...633..630W}
Willott C.~J.,  Delfosse X.,  Forveille T.,  Delorme P.,    Gwyn S.~D.~J.,
  2005, \apj, 633, 630

\bibitem[\protect\citeauthoryear{{Willott}, {Delorme}, {Reyl{\'e}}
  et~al.,}{{Willott} et~al.}{2010}]{2010AJ....139..906W}
{Willott} C.~J.,  {Delorme} P.,  {Reyl{\'e}} C.,    et~al., 2010, \aj, 139, 906

\bibitem[\protect\citeauthoryear{{Y{\`e}che}, {Petitjean}, {Rich}, {Aubourg},
  {Busca}, {Hamilton}, {Le Goff}, {Paris}, et~al.}{Y{\`e}che et~al.}{2010}]{2010A&A...523A..14Y}
{Y{\`e}che} C.,  {Petitjean} P.,  {Rich} J., et al.,  2010, \aap, 523, A14

\bibitem[\protect\citeauthoryear{{York}, {Adelman}, {Anderson} Jr.
  et~al.,}{{York} et~al.}{2000}]{2000AJ....120.1579Y}
{York} D.~G.,  {Adelman} J.,  {Anderson} Jr. J.~E.,    et~al., 2000, \aj, 120,
  1579

\bibitem[\protect\citeauthoryear{{Zhang}, {Pokorny}, {Jones} et~al.,}{{Zhang}
  et~al.}{2009}]{2009A&A...497..619Z}
{Zhang} Z.~H.,  {Pokorny} R.~S.,  {Jones} H.~R.~A.,    et~al., 2009, \aap, 497,
  619

\end{thebibliography}
\end{document}